\documentclass{article}

\usepackage[english]{babel}
\usepackage[a4paper,top=2cm,bottom=2cm,left=3cm,right=3cm,marginparwidth=2cm]{geometry}

\usepackage{amsmath, amssymb, amsthm}
\usepackage{mathtools}
\usepackage{graphicx}
\usepackage{siunitx}
\usepackage{hyperref}
\hypersetup{pdfauthor={Torben Schiz,Henrik Ebel},pdftitle={Reducing the Communication of Distributed Model Predictive Control: Autoencoders and Formation Control},colorlinks=false, hidelinks}
\usepackage{bm}
\usepackage{caption}
\usepackage[font=normalsize]{subcaption}
\usepackage{cite}
\usepackage{tikz}
\usetikzlibrary{calc, shapes, backgrounds, bending, decorations.pathmorphing, positioning, decorations.pathreplacing}
\usetikzlibrary{spy}
\usetikzlibrary{arrows.meta}
\usepgflibrary{arrows}
\usepackage{pgfplots}  
\usepackage{authblk}
\pgfplotsset{compat=newest}
\usepgfplotslibrary{colorbrewer}
\usepgfplotslibrary{polar}
\usepgfplotslibrary{statistics}
\usepackage{algorithmic, algorithm}
\newcommand*{\tran}{^{\mkern-1.5mu\mathsf{T}}} 
\usepackage{booktabs} 
\DeclareSIUnit\radian{rad}
\title{Reducing the Communication of Distributed Model Predictive Control: Autoencoders and Formation Control}

\author[1]{Torben Schiz}
\author[2]{Henrik Ebel\thanks{The corresponding author is H.~Ebel. \textit{Email address: }\texttt{henrik.ebel@lut.fi}}}

\affil[1]{University of Stuttgart, Stuttgart, Germany}
\affil[2]{Department of Mechanical Engineering, LUT University, Lappeenranta, Finland}

\begin{document}

\def\largeplot{0.2\linewidth}
\def\mainLineWidth{1.5pt}
\def\refLineWidth{1pt}
\def\layersep{2cm}
\definecolor{mycolor1}{HTML}{e41a1c}%
\definecolor{mycolor2}{HTML}{377eb8}%
\definecolor{mycolor3}{HTML}{4daf4a}%
\definecolor{mycolor4}{HTML}{ff7f00}%
\definecolor{mycolor5}{HTML}{ffff33}%
\definecolor{mycolor6}{HTML}{666666}%
\definecolor{mycolor7}{HTML}{a6cee3}%

\pgfplotsset{
	Myline1/.style={
		line width = \mainLineWidth, 
		mark=none,
		unbounded coords=jump,
		color=mycolor1
	}
}
\pgfplotsset{
	Myline2/.style={
		line width = \mainLineWidth, 
		mark=none,
		unbounded coords=jump,
		color=mycolor2
	}
}
\pgfplotsset{
	Myline3/.style={
		line width = \mainLineWidth, 
		mark=none,
		unbounded coords=jump,
		color=mycolor3
	}
}
\pgfplotsset{
	Myline4/.style={
		line width = \mainLineWidth, 
		mark=none,
		unbounded coords=jump,
		color=mycolor4
	}
}
\pgfplotsset{
	Myline5/.style={
		line width = \mainLineWidth, 
		mark=none,
		unbounded coords=jump,
		color=mycolor5
	}
}
\pgfplotsset{
	Myline6/.style={
		line width = \mainLineWidth, 
		mark=none,
		unbounded coords=jump,
		color=mycolor6
	}
}
\pgfplotsset{
	Myref1/.style={
		line width = \refLineWidth, 
		color=gray,
		dotted,
	}
}
\pgfplotsset{
	Myref2/.style={
		line width = \refLineWidth, 
		color=gray,
		dash dot,
	}
}
\pgfplotsset{
	Myref3/.style={
		line width = \refLineWidth, 
		color=gray,
		dashed,
	}
}
\pgfplotsset{
	MyStates1/.style={
	width=400pt,
	height=150pt,
	xlabel={Time in s},
	ylabel={States},
	}
}
\pgfplotsset{
	MyInput1/.style={
		width=400pt,
		height=150pt,
		xlabel={Time in s},
		ylabel={Inputs},
	}
}
\pgfplotsset{
	MyStates2/.style={
		width=210pt,
		height=150pt,
		xlabel={Time in s},
		ylabel={States},
	}
}
\pgfplotsset{
	MyInput2/.style={
		width=210pt,
		height=150pt,
		xlabel={Time in s},
		ylabel={Inputs},
	}
}
\pgfplotsset{
	MyTrajectory/.style={
		width=300pt,
		height=300pt,
		xlabel={$x$ in m},
		ylabel={$y$ in m},
	}
}
\date{}
\maketitle

\begin{abstract}
Communication remains a key factor limiting the applicability of distributed model predictive control (DMPC) in realistic settings, despite advances in wireless communication. 
DMPC schemes can require an overwhelming amount of information exchange between agents as the amount of data depends on the length of the predication horizon, for which some applications require a significant length to formally guarantee nominal asymptotic stability.
This work aims to provide an approach to reduce the communication effort of DMPC by reducing the size of the communicated data between agents. 
Using an autoencoder, the communicated data is reduced by the encoder part of the autoencoder prior to communication and reconstructed by the decoder part upon reception within the distributed optimization algorithm that constitutes the DMPC scheme. 
The choice of a learning-based reduction method is motivated by structure inherent to the data, which results from the data's connection to solutions of optimal control problems. 
The approach is implemented and tested at the example of formation control of differential-drive robots, which is challenging for optimization-based control due to the robots' nonholonomic constraints, and which is interesting due to the practical importance of mobile robotics. 
The applicability of the proposed approach is presented first in the form of a simulative analysis showing that the resulting control performance yields a satisfactory accuracy. 
In particular, the proposed approach outperforms the canonical naive way to reduce communication by reducing the length of the prediction horizon. 
Moreover, it is shown that numerical experiments conducted on embedded computation hardware, with real distributed computation and wireless communication, work well with the proposed way of reducing communication even in practical scenarios in which full communication fails, as the full-size data messages are not communicated in a timely-enough manner. 
This shows an objective benefit of using the proposed communication reduction in practice. 
\end{abstract}

\section{Introduction}
As the drive for automation continues, and as wireless communication, miniaturized computing, and sensors become more and more ubiquitous at low cost, the cooperative control of networked systems that shall fulfill a common task comes more and more into view. 
Collaborative control of power generators, power storage, and consumer demands can help stabilize power grids otherwise overwhelmed. 
Swarms of collaborating drones and robots can make much heavier machines operating individually seem obsolete in more and more applications. 
However, realizing this vast potential requires powerful distributed cooperative control methods. 
In particular, for widespread adoption, a desirable distributed control method should be widely applicable, including for nonlinear multiple-input multiple-output systems and systems with actuation constraints. 
At the same time, it should be easy to tune and configure for different goals, e.g., by specifying the cooperative goal in the form of a cost function. 
In general, distributed model predictive control~(DMPC), a distributed variant of the widely successful model predictive control~(MPC)~\cite{rawlings2017model}, fits the aforementioned requirements well and can even deal with goals not traditionally considered in asymptotically stabilizing or tracking control, as exemplified by economic (distributed) MPC methods~\cite{ferramosca2010EMPC, chen2012distributedEMPC, touretzky2014integratingEMPC}. 
In DMPC, each collaborating system solves in each time step its own optimal control(s) problem(s) using communicated information from neighboring systems, i.e., from those systems whose behavior influences the optimality of a system's own choices. 
However, two prices are paid for this. 
Firstly, one needs sufficient computation power to solve one or multiple optimal control problems numerically in each sampling step. 
This has been alleviated by ever faster computing devices, even in mobile and embedded form. 
Secondly, DMPC requires a quite strong communication network, as each solution (iterate) of a cooperative optimal control problem~(OCP) typically requires the exchange of candidate solutions over the whole prediction horizon over which the model predictive controller predicts the system performance. 
Whereas better and faster communication technologies, like 5G networking, have helped for that, communication still represents a noteworthy limitation in many practical applications, as even the most advanced communication technology can be impacted by disturbances or a crowded electromagnetic frequency spectrum, limiting throughput. 
Indeed, even in laboratory settings and very recent work~\cite{stomberg2024cooperative}, it has been seen that communication is typically a key limiting factor for the application of DMPC when computation really happens in a physically distributed fashion with wireless communication. 

This article contributes a method to successfully reduce the communication footprint of DMPC while still attaining a sufficient control accuracy. 
We demonstrate this in a nonlinear control setting, namely the formation control of nonholonomic mobile robots. 
We pick this setting as a model problem for various reasons. 
Firstly, a generic task in cooperative distributed control is that systems coordinate their states or outputs relative to one another and, potentially, also to an absolute reference. 
For mobile robots, outputs of interest are usually the robots' poses, making formation control a physical embodiment of the abstract task of distributed state or output coordination. 
Secondly, it is known that the asymptotic setpoint stabilization of nonholonomic robots is challenging for optimal control~\cite{muller2017quadratic, rosenfelder2021cooperative, rosenfelder2023automatica, ebel2023PAMM}. 
Thirdly, mobile robots and swarms of mobile robots are of increasing practical importance, in transportation and logistics, in security and defence, in service robotics, etc.
Consequently, communication-based cooperation is expected to increase productivity and may revolutionize applications such as logistics~\cite{grau2017industrial,wen2018swarm}.
Specifically, formation control of mobile robots is seen as an important task in distributed robotics~\cite{bai2011cooperative, olfati2007consensus}.

In idealized laboratory settings, DMPC has already yielded impressive results for instance in collaborative robotics ~\cite{burk2021experimental, ebel2021distributed, rosenfelder2022, stomberg2023cooperativecompare, stomberg2024cooperative, van2017distributed, ebel2024coop_ob_trans, tavares2020dynamic, pauca2024dmpc}.
Interestingly, although it has been recognized that communication is a key factor limiting DMPC's practical applicability and performance~\cite{stomberg2024cooperative}, there is, to the best knowledge of the authors, barely any research on reducing DMPC's communication demands. 
One root cause may be that, generally, there are very little works that actually apply DMPC on physically distributed hardware, with real communication and one computer per control agent. 
The only lines of work that the authors are aware of that use DMPC in more or less realistic settings in robotics are represented by~\cite{van2017distributed, stomberg2023cooperativecompare, stomberg2024cooperative, pauca2024dmpc}. 
Thereof, the only work considering the most realistic setup with real on-board computation and wireless communication  is~\cite{stomberg2024cooperative}, which, however, identifies abundant communication as a key limiting factor. 

Thus, as so few works consider realistic settings, one may argue that there might have been little research interest to reduce communication. 
It is worth pointing out here that, thinking of highly dynamic robotics applications, we are not interested in intermittent communication, where communication is reduced by communicating as seldomly as possible (which can be attractive, e.g., in an internet-of-things setting where dynamics are often slow). 
Thus, we do not want to send fewer messages, but we intend to reduce the amount of data communicated per message. 
The only works the authors are aware of that try to reduce communication this way for DMPC are~\cite{elferik2013DNMPC, elferik2015distributed}.
Both works deal with distributed nonlinear MPC and communicate the parameters of a small neural network. 
Using the neural network described by these parameters, the receiving agent reconstructs the predicted trajectories of the sending agent.
By communicating the weights and biases of the neural network, the amount of floating-point data sent per message is reduced significantly.
However, this approach requires the retraining of a neural network in each time step, which causes a considerable computational overhead.
In consequence, it is noted in~\cite{elferik2015distributed} that it takes \qty{94}{\second} of CPU time to simulate one second, despite parallel commutation being used. 
Thus, the approach is far from meeting any realistic real-time requirements.

Hence, to summarize the state of the art, to the best knowledge of the authors, the problem tackled in this work has been barely studied so far.
The key novel contribution of this work is a real-time capable communication reduction for DMPC using a learning-based method.
For this, a pre-trained, standard autoencoder~\cite{lecun1987, bourlard1988auto, hinton1993autoencoders} is built into the distributed optimization algorithm from~\cite{stewart2011} as, e.g., useful for formation control of mobile robots~\cite{rosenfelder2022}.
Messages are encoded prior to transmission by one agent and decoded immediately after reception by another.
Although both DMPC and autoencoders are known techniques, they are, to the authors' knowledge, combined here for the first time in an effort to meet real-world requirements in DMPC by compressing data packages. 
The usage of a learning-based approach to reducing the communication is motivated by inherent structure in the communicated data.
This structure in the data stems from its connection to the nonlinear system dynamics and MPC's underlying receding-horizon principle as it is closely related to the controller's prediction of the system's future behavior. 

This work is structured as follows. We begin with an overview of the problem setup and general approach of this work in Section~\ref{sec:problem_setup}.
In Section~\ref{sec:theory}, we recapitulate the work's methodological foundations, focusing on the solution of DMPC problems and how DMPC can be used for the formation control of nonholonomic robots. 
Section~\ref{sec:add_ae_to_opti_algo} is dedicated to the training and hyperparameter tuning process of the autoencoder.
The results are presented in Section~\ref{sec:results}.
The results include a simulative analysis of the reduced communication, starting with an idealized setting to study purely the error resulting from the reduced communication, and continuing with a more realistic setup with plant-model mismatch. 
We will see that this work's proposed way of compressing the communicated data performs better than the canonical naive approach to simply reduce the number of steps in the prediction horizon.
Finally, the communication reduction is employed in a physically distributed setting on computation hardware as it is typical for mobile robots to analyze the real-world applicability of the presented approach.
Section~\ref{sec:conclusion} provides a concluding summary of the findings of the work.

\section{Problem Setup and General Approach}\label{sec:problem_setup}
\begin{figure}[btp]
     \centering
    \includegraphics{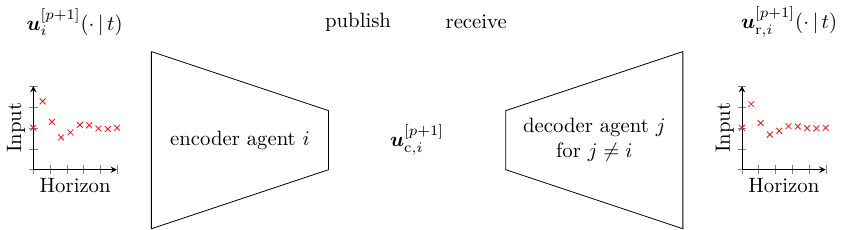}
    \caption{Autoencoder as built into the distributed optimization algorithm.}
    \label{fig:split_autoencoder}
\end{figure}

Subsequently, we consider a set of~$N\geq 2$ dynamically decoupled control systems that shall pursue a common goal as described by a cost function $J$ coupling the states or outputs of the individual control systems. 
More concretely, we study the case where this cooperative control problem is formulated as a model predictive control problem of the form 
\begin{align}
\underset{\bm u(\cdot \,\vert \, t )}{\textnormal{minimize}} & & & J(\bm x(t), \bm u(\cdot \,\vert \, t ))\label{eq:mpccost}\\
\textnormal{subject to} & & &\bm x(t+k+1\,\vert\, t) = \bm f(\bm x(t+k\,\vert\, t), \bm u(t+k\,\vert\, t)), \label{eq:discretizationofsysdym}\\
&&& \bm u(t+k\,\vert\, t) \in \mathcal{U}=\mathcal{U}_1\times\dots\times\mathcal{U}_N, \, k \in \{0,1\dots,H-1\},\\
&&& \bm x(t\,\vert\, t) = \bm x(t), \label{eq:mpcinit}
\end{align}
where~\eqref{eq:discretizationofsysdym} is a discretization of the concatenated dynamics of the individual control systems' independent dynamics with $\bm x\in\mathbb R^{n_x}$, and the set~$\mathcal{U}$ results from independent, per-system constraints on the control inputs~$\bm u\in\mathbb R^{n_u}$ of the control systems.
Here, the cost function is defined as $J(\bm x(t), \bm u(\cdot \,\vert \, t ))=\sum_{k=0}^{H-1} \ell (\bm x(t+k\,\vert\, t), \bm u(t+k\,\vert\, t))$, where~$\ell\!:\mathbb R^{n_x} \times \mathcal U \to \mathbb R$ is the stage cost and $H\in\mathbb N$ is the length of a finite prediction horizon.
For MPC-related variables, a trajectory predicted at time $t$ along the horizon of length $H$ is denoted as $(\cdot\, \vert\, t)$.
If we refer to the $k$th value in the prediction horizon, we write $(t+k\,\vert\, t)$ with $k \in \{0,\dots, H-1\}$. 
In the concrete example studied, the control systems will be nonholonomic mobile robots described by their nonlinear first-order kinematics, and their cooperative control goal will be formation control. 
We assume that each control system shall make its own decisions, relying on communicated information from those control systems that are relevant to the optimality of the system's respective control decisions. 
Depending on the precise properties of the control systems and the cost function, there exist many distributed model predictive control techniques that achieve this, see, e.g.,~\cite{stewartvenkat2010cooperative, stewart2011, mullerreble2012cooperative}. 
In all of these, for each control system, at least one optimal control problem is solved per time step, and iterative schemes can have multiple iterations per time step to come closer to a hypothetical centralized solution. 
What all established methods have in common is that, in each time step, they require the exchange of data between the subsystems, where the size of the data typically scales linearly with the prediction horizon~$H$. 
Commonly, the communicated data consists of candidate solutions of the individual optimization problems being solved. 
Depending on the precise solution algorithm, variants are conceivable (e.g., sending incremental updates of the candidate solutions instead), but the size dependency on~$H$ and the general relationships to an optimal solution of an OCP and to the underlying system dynamics persist. 
Thus, instead of sending the raw data, spanning each time step in the prediction horizon, in this work, we intend to learn a more efficient data representation using an undercomplete autoencoder artificial neural network~\cite{lecun1987, bourlard1988auto, hinton1993autoencoders}, of which the principle is schematically depicted in Figure~\ref{fig:split_autoencoder} for control system~$i\in\mathbb{N}$. 
Therein, without loss of generality, it is assumed that system~$i$ communicates $\bm u_i^{[p+1]}(\cdot \, \vert \, t)$, a candidate control-input trajectory at time step~$t$ and solver iteration~$p+1$, of which $\bm u_{\textnormal{c},i}^{[p+1]}$ represents a compressed version in the latent space. 
Then, $\bm u_{\textnormal{r},i}^{[p+1]}(\cdot \, \vert \, t)$ represents a recovered version that, ideally, would be identical to $\bm u_i^{[p+1]}(\cdot \, \vert \, t)$. 
The goal is that $\bm u_{\textnormal{c},i}^{[p+1]}$ is considerably smaller (in terms of memory) than the original data so that, by sending $\bm u_{\textnormal{c},i}^{[p+1]}$, the communication network is occupied less. 
At the same time, the error between the recovered data $\bm u_{\textnormal{r},i}^{[p+1]}(\cdot \, \vert \, t)$ and the original data must be small enough so that closed-loop performance still meets the needs of the application in question. 

All implementations in this work use a fully distributed software structure for training-data generation and all numerical experiments. 
In particular, the control logic of each control system always runs in its own program, implemented in Python, and data is exchanged via the LCM package~\cite{huang2010lcm} using UDP multicast network messaging. 
Moreover, in simulation scenarios, the physics is simulated in a separate simulator program, which interfaces also via LCM messages. 
This kind of setup ensures that distributed computation and communication are always realistically considered, and it can work without software changes both in networked simulations and with robot hardware, see~\cite{ebel2022cooperative}. 
For training-data collection, a separate data recorder, implemented in a multi-threaded fashion in C++ and also receiving messages via LCM, is used to record the required data from inter-agent communication as reliably as possible. 

Due to the subject matter of this work, fundamentals from two main directions are required, as discussed subsequently.

\section{Methodological Foundations}\label{sec:theory}
The distributed MPC approach used in this work, including the proposed way of reducing communication, are not limited to a very specific cooperative control task. 
Thus, they can be readily used also in tasks other than formation control and mobile robotics. 
Despite that, in Section~\ref{sec:nonholonomic_robot_and_formation}, we first recapitulate based on~\cite{rosenfelder2022} how formation control with (nonholonomic) differential-drive robots can be furnished using DMPC, as it will be used later in the case studies.  
Introducing the required concepts and notation directly at the example of formation control makes the concepts more descriptive and understandable. 
In Section~\ref{sec:distributed_algo}, we concisely introduce the distributed solution algorithm from~\cite{stewart2011}, which we use in this work as it can deal with nonlinear dynamics and as it yielded good results in our previous work~\cite{rosenfelder2022}. 
Focal point is the information necessary to understand how the algorithm can be modified to account for the reduced communication.
\subsection{Nonholonomic Mobile Robots and Formations}\label{sec:nonholonomic_robot_and_formation}

\begin{figure}
    \centering
    \includegraphics[width=0.3\linewidth]{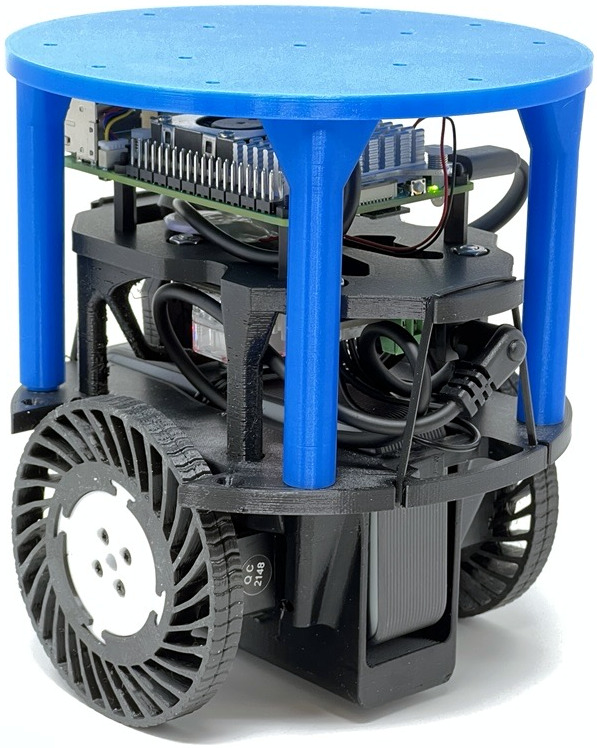}
    \caption{Exemplary differential-drive mobile robot}
    \label{fig:mobilerobot}
\end{figure}

This study considers differential-drive mobile robots due to their popularity in service robotics and their challenges to control because of their nonholonomic kinematic constraint.
A differential-drive robot has two individually driven wheels on a common axis and a freely turning caster wheel or ball.
An exemplary version of such a robot is depicted in Figure~\ref{fig:mobilerobot}.
The robot moves in a horizontal plane. 
It is assumed to roll without slipping and thus subject to a nonholonomic constraint.
The continuous-time nonlinear model
\begin{align}
    \dot{\bm z}_i(t) = \begin{bmatrix}
        \dot x_i (t) \\ \dot y_i (t) \\ \dot\theta_i (t) 
    \end{bmatrix} = \begin{bmatrix}
        v_i(t) \cos(\theta_i(t)) \\ v_i(t) \sin(\theta_i(t)) \\ \omega_i (t)
    \end{bmatrix} \eqqcolon \bm g_1(\bm z_i(t))v_i(t) + \bm g_2(\bm z_i(t))\omega_i(t), \; \bm z_i(0) = \bm z_{i,0}
\end{align}
with the state vector ${\bm z}_i(t) = \begin{bmatrix}x_i (t) &  y_i (t) & \theta_i (t))\end{bmatrix}\tran \in \mathbb R^{n_i}$, $n_i=3$, describes the first-order kinematics of such a nonholonomic robot.
The state vector incorporates the spatial position $\begin{bmatrix} x_i & y_i \end{bmatrix}\tran$ in the inertial frame of reference and the orientation angle $\theta_i$ measured from the positive $x_i$-axis.
The robot's kinematics is subject to the Pfaffian constraint $\begin{bmatrix}\sin \theta & -\cos\theta_i & 0\end{bmatrix}\dot{\bm z}_i = 0$, which prevents instantaneous lateral motions of the robot \cite[p. 298]{luca1995modelling}.
The linear velocity $v_i$ and the angular velocity $\omega_i$ form the control input $\bm u_i = \begin{bmatrix} v_i & \omega_i \end{bmatrix}^\top$.
The control input is subject to pointwise-in-time constraints $\bm u_i \in \mathcal U_i$, restricted here to $\bm u_i \in \mathcal U_i = \left[\underline{v_i}, \overline{v_i}\right] \times \left[\underline{\omega}_i, \overline{\omega}_i\right]\subset \mathbb R^{m_i}$ where $m_i = 2$ and $\underline \bullet$ and $\overline \bullet$ represent lower and upper limits, respectively.

The system is driftless and completely controllable~\cite{rosenfelder2023automatica,rosenfelder2022}. 
Characteristically, when using MPC without terminal ingredients, even a single differential-drive robot cannot be asymptotically stabilized with an, in other applications prevalent and almost uniquely employed, quadratic cost function. 
Instead, following~\cite{worthmann2015regulation}, asymptotic stability of the MPC closed loop can be achieved using a non-quadratic cost function~\cite{rosenfelder2022}, which represents a better approximation of distance for the system's underlying sub-Riemannian geometry.
A characteristic feature of the stage cost is that deviations from the desired state in the $y_i$-direction enter the cost with an exponent of \num{2} whereas the other components enter with an exponent of \num{4}. 
Due to this structure of the stage cost, close to the origin, the main aim of the optimization is to reduce the error in the $y_i$-direction.
A theoretical discussion of the closed-loop stability of MPC controllers designed this way for nonholonomic systems can be found in~\cite{muller2017quadratic, rosenfelder2023automatica}.

To accurately describe the control task, usually given in an inertial frame of reference, and quantities as perceived by the robots and their components, various frames of reference are utilized. 
Subsequently, if a variable $v$ is described in a specific coordinate frame $\mathcal K_\textnormal{C}$, the corresponding superscript is added to emphasize this, i.e., $v^\textnormal{C}$.
The inertial frame of reference is $\mathcal K_\textnormal{I}$ and a variable $v$ described in it is denoted without a superscript, i.e., $v \coloneqq v^\textnormal{I}$.
We consider only right-handed frames of reference such that one frame of reference only differs from another by a rotation around the $z$-axis.

Considering a formation of mobile robots, we can write the overall state $\bm z = \begin{bmatrix}
 \bm z_1\tran & \dots & \bm z_N\tran   
\end{bmatrix}\tran \in \mathbb R^n$, $n=3N$ and input $\bm u = \begin{bmatrix}
 \bm u_1\tran & \dots & \bm u_N\tran   
\end{bmatrix}\tran \in \mathbb R^m$, $m=2N$, by concatenating the states $\bm z_j$ and inputs $\bm u_j$, $j\in\mathcal N \coloneqq \mathbb Z_{1:N}$ of all robots as they are dynamically decoupled, yielding the overall dynamics
\begin{align}
    \dot{\bm z} (t) = \begin{bmatrix}
        \dot{\bm z}_1 (t) \\ \vdots \\ \dot{\bm z}_N (t)   
    \end{bmatrix} = \begin{bmatrix}
        \bm G({\bm z}_1 (t)) & & \\ & \ddots & \\ & & \bm G({\bm z}_N (t))
    \end{bmatrix}\begin{bmatrix}
        \bm u_1 (t) \\ \vdots \\ \bm u_N (t)
    \end{bmatrix}  \eqqcolon \bm G_{\textnormal{c}} (\bm z (t)) \, \bm u (t), \; \bm z(0) = \bm z_0,
\end{align}
where $\bm G (\bm z_j(t)) = \begin{bmatrix}\bm g_1 (\bm z_j(t)) & \bm g_2(\bm z_j(t))\end{bmatrix} \in \mathbb R^{n_i \times m_i}$ and, hence, $\bm G_{\textnormal{c}} \, : \, \mathbb R^n \to \mathbb R^{n\times m}$.
The discretized system dynamics, using a time-discretization with zero-order hold on the control inputs and sampling time $\delta t$, reads $\bm z(t + \delta t) = \bm G_{\textnormal{d}} (\bm z(t), \bm u(t))$. 
Choosing an output $\bm y = \bm C \bm z$ allows to describe the robot formation.
We choose an output that includes the geometric center of the formation as well as the positions of each robot relative to the geometric center, where $\begin{bmatrix} x_{\hat {\bm z} \to j,\textnormal{d}} & y_{\hat {\bm z} \to j,\textnormal{d}} \end{bmatrix}\tran$ denotes the relative position of robot~$j$.
This allows penalizing the geometric center and the relative positions individually via the weights in the cost function.
The geometric center acts as a virtual leader of the formation.
Furthermore, this enables a balance between maintaining relative positions within the formation and fast movement of the formation to its goal position.
The notation $\hat{\bullet}$ is used to denote any variable referring to the geometric center, e.g., $\hat {\bm z}$ is the geometric center's pose. 
In order to allow each robot and the virtual leader to approach the setpoint along the $x^\textnormal{R}$-axis of an auxiliary reference frame, we only consider scenarios in which each robot has a desired orientation of zero relative to the virtual leader.
The corresponding reference frame $\mathcal K_\textnormal{R}$ originates from a rotation of the inertial frame of reference by the desired orientation $\hat\theta_{\textnormal{d}}$ of the geometric center.
All together, this leads to the output
\begin{align}\label{eq:outputvector}
    \bm y(\bm z, \hat \theta_d) = \begin{bmatrix}
        \hat x^\textnormal{R} \\ \hat y^\textnormal{R} \\ \hat \theta \\ x_{\hat{\bm z}\to 1}^\textnormal{R} \\ y_{\hat{\bm z}\to 1}^\textnormal{R} \\ \theta_1 \\ \vdots \\ x_{\hat{\bm z}\to N}^\textnormal{R} \\ y_{\hat{\bm z}\to N}^\textnormal{R} \\ \theta_N
    \end{bmatrix} = \begin{bmatrix}
        \sum_{i \in \mathcal N} \frac1N x_i^\textnormal{R} \\ \sum_{i \in \mathcal N} \frac1N y_i^\textnormal{R} \\
        \sum_{i \in \mathcal N} \frac1N \theta_i \\
        x_1^\textnormal{R} - \sum_{i \in \mathcal N} \frac1N x_i^\textnormal{R} \\
        y_1^\textnormal{R} - \sum_{i \in \mathcal N} \frac1N y_i^\textnormal{R} \\
        \theta_1 \\ \vdots \\
        x_N^\textnormal{R} - \sum_{i \in \mathcal N} \frac1N x_N^\textnormal{R} \\
        y_N^\textnormal{R} - \sum_{i \in \mathcal N} \frac1N y_N^\textnormal{R} \\
        \theta_N
    \end{bmatrix} \in \mathbb R^{3(N+1)},
\end{align}
consisting of the geometric center of the formation, the position of each robot relative to the geometric center, and the robots' orientations.
The relative position of each robot $i\in\mathcal N$ is given in the frame of reference $\mathcal K_\textnormal{R}$ by rotation in the plane about the angle $\hat \theta_\textnormal{d}$ by
\begin{align}
   \begin{bmatrix}
       x_i^\textnormal{R} \\ y_i^\textnormal{R} 
   \end{bmatrix} = \begin{bmatrix}
           \cos \hat\theta_\textnormal{d}  & \sin\hat\theta_\textnormal{d} \\ -\sin\hat\theta_\textnormal{d} & \cos \hat\theta_\textnormal{d}
       \end{bmatrix} \begin{bmatrix}
           x_i \\ y_i
       \end{bmatrix} .
\end{align}
The output $\bm y^\textnormal{R}(\bm z, \hat \theta_\textnormal{d})$ can be rewritten as $\bm y^\textnormal{R}(\bm z, \hat \theta_\textnormal{d}) = \bm C^\textnormal{R}(\hat\theta_\textnormal{d})\bm z$ using an output matrix $\bm C^\textnormal{R}(\hat\theta_\textnormal{d})$.
Applying the output matrix to the overall state vector $\bm z$ rotates the positions by $\hat \theta_{\mathrm{d}}$, adds the position of the geometric center by calculating the mean of the absolute state of each robot, and formulates the position of each robot relative to the geometric center resulting in~\eqref{eq:outputvector}.
For a more detailed description of the reformulation, we refer to~\cite{rosenfelder2022}.
This allows each robot and the virtual leader to approach any setpoint along the $x^\textnormal{R}$-axis of an auxiliary reference frame.

We restrict the description in this work to the setpoint stabilization problem and express the goal of the formation task by means of the desired output
\begin{align}
    \bm y_\textnormal{d}^\textnormal{R} = \begin{bmatrix}\hat x_\textnormal{d}^\textnormal{R} & \hat y_\textnormal{d}^\textnormal{R} & \hat \theta_\textnormal{d} & x_{\hat {\bm z} \to 1,\textnormal{d}} & y_{\hat {\bm z} \to 1,\textnormal{d}} & \theta_{1,\textnormal{d}} & \cdots &  x_{\hat {\bm z} \to N,\textnormal{d}} & y_{\hat {\bm z} \to N,\textnormal{d}} & \theta_{N,\textnormal{d}}  \end{bmatrix}\tran \in \mathbb R^{3(N+1)},
\end{align}
with $\theta_{j,\textnormal{d}} = \hat \theta_{\textnormal{d}}$ for all $j \in \mathcal N$.
Due to the redundant formulation of the output $\bm y^\textnormal{R}(\bm z, \hat \theta_\textnormal{d})$, which is used to treat all robots equally, a certain consistency condition must hold. 
The relative positions of the robots must satisfy 
\begin{align}\label{eq:consistency}
    \sum_{j \in \mathcal N} \begin{bmatrix} x_{\hat {\bm z} \to j,{\textnormal{d}}} & y_{\hat {\bm z} \to j,{\textnormal{d}}} \end{bmatrix}\tran = \bm 0 \;\text{as well as} \; \theta_{k,\textnormal{d}} = \hat \theta_{\textnormal{d}} \, \text{for all} \, k \in \mathcal N,
\end{align}
which can be readily checked when defining the control task. 
Reformulating the stage cost for a single robot for the formation task results in 
\begin{align}\label{eq:formation_stage_cost}
    \ell (\bm z, \bm u, \bm y_{\textnormal{d}}^\textnormal{R}) = \hat\ell (\bm z, \bm u, \bm y_{\textnormal{d}}^\textnormal{R}) + \sum_{j = 1}^N \left(\ell_{j,\textnormal{rel}} \left(\bm z, \bm u, \bm y_{\textnormal{d}}^\textnormal{R}\right) + \ell_{j,u} (\bm u)\right),
\end{align}
where
\begin{align}
    \hat \ell (\cdot) &\coloneqq \hat d_1\left(\hat x^\textnormal{R} - \hat x_{\textnormal{d}}^\textnormal{R}\right)^4 + \hat d_2\left(\hat y^\textnormal{R} - \hat y_{\textnormal{d}}^\textnormal{R}\right)^2 + \hat d_3\left(\hat \theta^\textnormal{R} - \hat \theta_{\textnormal{d}}^\textnormal{R}\right)^4, \\
    \hat \ell_{j,\text{rel}} (\cdot) &\coloneqq \hat d_{1,j}\left(\hat x_{\hat {\bm z} \to j}^\textnormal{R} - \hat x_{\hat {\bm z} \to j,{\textnormal{d}}}^\textnormal{R}\right)^4 + \hat d_{2, j}\left(\hat y_{\hat {\bm z} \to j}^\textnormal{R} - \hat y_{\hat {\bm z} \to j,{\textnormal{d}}}^\textnormal{R}\right)^2 + \hat d_{3, j}\left(\hat \theta_j^\textnormal{R} - \hat \theta_{j,{\textnormal{d}}}^\textnormal{R}\right)^4, \\
    \ell_{j,u} (\cdot) &= r_{1,j} v_j^4 + r_{2,j} \omega_j^4
\end{align}
with the weights $\hat d_i, \, d_{i,j}, \, r_{m,j} \in \mathbb R_{>0}$, $i \in \mathbb Z_{1:3}$, $j \in \mathcal N$, $m \in \mathbb Z_{1:2}$.
Here and in the following, we denote a set of integers as $\mathbb Z_{a:b} = \{a, a+1, \dots, b -1, b\}$  given the bounding integers $a \leq b$.

Summing the stage cost~\eqref{eq:formation_stage_cost} over the prediction horizon gives the cost function 
\begin{align}\label{eq:formation_cost}
    J\!\left(\bm z(t), \bm u(\cdot \, \vert \, t), \bm y_{\textnormal{d}}^\textnormal{R}\right) = \sum_{k = 0}^{H - 1} \ell(\bm z(t + k \,\vert\, t), \bm u(t + k \,\vert\, t) ,\bm y_{\textnormal{d}}^\textnormal{R}).
\end{align}
Each robot optimizes only its own control inputs instead of the overall control input.
This leads to a cooperative distributed optimization problem.
We formulate in the following, without loss of generality, the OCP for robot $\hat\imath \in \mathcal N$ with the state $\bm z_{\hat\imath}$ and the control input $\bm u_{\hat\imath}$.
To distinguish between the input $\bm u_{\hat\imath}$ of robot $\hat\imath$ and the input of the other robots, which are collected in $\bm{u}_{-\hat\imath} \coloneqq \begin{bmatrix}\bm u_{1}\tran & \cdots & \bm u_{\hat\imath - 1}\tran & \cdots & \bm u_{\hat\imath + 1}\tran & \cdots & \bm u_{N}\tran \end{bmatrix}\tran \in \mathbb R^{m - m_i}$, we write the overall control input as $\bm u(\bm u_{\hat\imath}, \bm u_{-\hat\imath})$. 

Using the cost~\eqref{eq:formation_cost} and the discretized system dynamics, the DMPC optimization problem for robot $\hat\imath$ at time $t$ takes the form
\begin{align}
    &\underset{\bm u_{\hat\imath}(\cdot \, \vert \, t)}{\text{minimize}} \; && J\!\left(\bm z(t), \bm u(\bm u_{\hat\imath}(\cdot \, \vert \, t), \bm u_{-\hat\imath}(\cdot \, \vert \, t)), \bm y_\textnormal{d}^\textnormal{R}\right)\label{eq:dist_cost} \\
    &\text{subject to} \; && \bm z(t + k + 1 \,\vert\, t) = \bm G_\textnormal{d} \left(\bm z(t + k\,\vert \, t), \bm u(\bm u_{\hat\imath}(t + k \,\vert \, t), \bm u_{-\hat\imath}(t + k \,\vert\, t))\right), \label{eq:dist_sys_dyn}  \\
    & && \bm u(t + k \,\vert\, t) \in \mathcal U, \, k \in \mathbb Z_{0:H-1}, \label{eq:inputconstraints} \\
    & && \bm z(t \,\vert\, t) = \bm z(t). \label{eq:initial_state_is_current}
\end{align}

\subsection{Distributed Optimization for Nonconvex Problems}\label{sec:distributed_algo}
As our method of choice to allow distributed optimization in the following sections, this section describes the distributed algorithm from~\cite{stewart2011} for iteratively solving the optimal control problem~\eqref{eq:dist_cost}-\eqref{eq:initial_state_is_current}. 
The algorithm from~\cite{stewart2011} ensures a non-increasing objective function as well as feasibility from iteration to iteration. 
By substituting the predicted states~\eqref{eq:dist_sys_dyn} into the cost function~\eqref{eq:dist_cost}, the problem becomes subject only to input constraints.
This can result in a nonconvex optimal control problem as a consequence of the nonlinearity of the system dynamics, yielding the objective function~$\tilde{J}$ with inserted system dynamics  at time $t$ and iteration $p$ as 
\begin{align}
    \min_{\bm u_{\hat \imath}(\cdot\, \vert\, t) \in \mathcal U_{\hat \imath}} \tilde{J}\!\left(\bm z(t), \bm u\!\left(\bm u_{\hat \imath}^{[p]} (\cdot\, \vert\, t), \bm u_{-\hat \imath}^{[p]}(\cdot\, \vert\, t) \right)\! , \bm y_{\textnormal{d}}^{\textnormal R}\right).
\end{align}
The initial candidate input for agent $\hat \imath$, which, in our case, will be a robot, is denoted as $\bm u_{\hat \imath}^{[0]} (\cdot \,\vert\, t)$ and the solution candidate at iteration $p$ as $\bm u_{\hat \imath}^{[p]} (\cdot \,\vert\, t)$.
The following iteration with iteration index $p+1$ is computed based on $\bm u\!\left(\bm u_{\hat \imath}^{[p]}, \bm u_{-\hat \imath}^{[p]}\right)$, the overall state $\bm z(t)$, and the desired setpoint $\bm y_\textnormal{d}^\textnormal{R}$.

In a first step, the algorithm uses a line search
\begin{align}\label{eq:projected_u}
    \bar{\bm u}_{\hat \imath}^{[p]} = \mathcal P\!\left(\bm u_{\hat \imath}^{[p]}(\cdot\, \vert\, t) - \nabla_{\hat \imath}\tilde{J}\!\left(\bm z(t), \bm u\! \left(\bm u_{\hat \imath}^{[p]} (\cdot\, \vert\, t), \bm u^{[p]}_{-\hat \imath}(\cdot\, \vert\, t) \right)\!, \bm y_{\textnormal{d}}^{\textnormal R}\right)\right)
\end{align}
to compute an approximate solution, where $\nabla_{\hat \imath}\tilde{J}$ is the $\hat \imath$th component of the objective function's gradient and the function $\mathcal P(\cdot)$ a projection onto the set $\mathcal U_{\hat \imath}$.
This projection ensures that the input constraints are not violated. 
Next, the step is determined by multiplying the step size $\alpha_{\hat \imath}^{[p]}$ to the step direction $\bm \nu_{\hat \imath}^{[p]} = \bar{\bm u}_{\hat \imath}^{[p]}(\cdot \, \vert \, t) - \bm u_{\hat \imath}^{[p]}(\cdot \, \vert \, t)$.
To determine a suitable value for the step size, the algorithm applies a backtracking line search: Starting from an initial step size $\bar \alpha_{\hat \imath}$, its value is iteratively decreased by multiplication with a backtracking factor $\beta\in (0,1)$, i.e., $\alpha_{\hat \imath}^{[q]} = \beta\alpha_{\hat \imath}^{[q - 1]}$ with  $\alpha_{\hat \imath}^{[0]} = \bar \alpha$.
In this work, each agent~$\hat \imath \in \mathcal N$ takes the same initial values for $\bar\alpha$ and $\beta$ as the agents are assumed to be equal in our robotics application later. 
We omit the iteration superscript $p$ from the iterative update of the step size for the sake of readability.
This shrinking process of the step size is repeated until it satisfies the Armijo rule~\cite[Proposition 2.3.3]{bertsekas1997nonlinear}
\begin{align}\label{eq:armijo}
\begin{split}
    \tilde{J}\!\left(\bm z(t), \bm u \!\left(\bm u_{\hat \imath}^{[p]} (\cdot\, \vert\, t), \bm u^{[p]}_{-\hat \imath}(\cdot \,\vert\, t) \right)\! , \bm y_{\textnormal{d}}^{\textnormal R}\right) -\tilde{J}\!\left(\bm z(t), \bm u \!\left(\bm u_{\hat \imath}^{[p]} (\cdot \,\vert\, t) + \alpha_{\hat \imath}^{[q]} \bm\nu_{\hat \imath}^{[p]}, \bm u^{[p]}_{-\hat \imath}(\cdot \,\vert\, t) \right)\!, \bm y_{\textnormal{d}}^{\textnormal R}\right) \\ \geq - \sigma \alpha_{\hat \imath}^{[q]}\nabla_{\hat \imath}\left[\tilde{J}\!\left(\bm z(t), \bm u \!\left(\bm u_{\hat \imath}^{[p]} (\cdot \,\vert\, t), \bm u^{[p]}_{-\hat \imath}(\cdot \,\vert\, t) \right)\! , \bm y_{\textnormal{d}}^{\textnormal R}\right)\right]\tran \bm\nu_{\hat \imath}^{[p]}    
\end{split}
\end{align}
with $\sigma \in (0,1)$. 
We denote the step size  for which the Armijo rule is first fulfilled by $\alpha_{\hat \imath}^{[p]}$.
Adding the weighted product of step size and step direction to $\bm u_{\hat \imath}^{[p]}$ yields a candidate input
\begin{align}\label{eq:candidate_input}
\bm u_{\hat \imath}^{[p+1]}(\cdot\, \vert\, t) = \bm u_{\hat \imath}^{[p]}(\cdot\, \vert\, t) + w_{\hat \imath} \alpha_{\hat \imath}^{[p]}\bm\nu_{\hat \imath}^{[p]}
\end{align}
with weight $w_{\hat \imath} > 0$ subject to $\sum_{j \in \mathcal N} w_j = 1$ and initialized as $w_j = 1/N$ for all $j\in \mathcal N$ for each subproblem.
At this stage in the optimization algorithm, the agents exchange their respective inputs.
To reduce the communication effort, this work introduces a modification to the algorithm by including the encoder part of an autoencoder to encode the candidate input $\bm u_{\hat \imath}^{[p+1]}(\cdot\, \vert\, t)$ to a reduced representation $\bm u_{\hat \imath, \textnormal{c}}^{[p+1]}$ before publishing the data.
As we aim to reduce the amount of communicated data, we solely consider undercomplete autoencoders, i.e., the input dimension is larger than the dimension of the encoded representation.
The agent publishes the reduced candidate inputs $\bm u_{\hat \imath, \textnormal{c}}^{[p+1]}$ and reconstructs the candidate inputs it receives from the other agents from the reduced representation by a forward pass through the decoder part of the autoencoder yielding the reconstructed candidate input sequence $\bm u_{j, \textnormal{r}}^{[p+1]}(\cdot\, \vert\, t)$ for $j\in\mathcal N \setminus\lbrace\hat{\imath}\rbrace$.
In other words, the autoencoder is split into its encoder and decoder part, which are applied before sending and immediately after receiving, respectively.
As each agent anyway always has access only to a reconstruction of the communicated data, we omit the reconstruction index $\textnormal{r}$ in the following for readability. 

We use the received current candidate inputs together with the candidate inputs from the previous iteration to compute the step for agent $\hat \imath$ as
\begin{align}\label{eq:step_from_u}
    \bm \gamma_j^{[p]} \coloneqq \alpha_j^{[p]} \bm \nu_j^{[p]} = \frac{\bm u_{j}^{[p + 1]}(\cdot\, \vert\, t) - \bm u_{j}^{[p]}(\cdot\, \vert\, t)}{\omega_j} \; \text{for} 
\end{align}
for $j \in \mathcal N\setminus\lbrace \hat \imath\rbrace$.

With these results, the algorithm enters the second part.
The agents check simultaneously if the cost function is convex-like to ensure a non-increasing objective function from iteration to iteration by checking if the inequality 
\begin{align}\label{eq:convex_like}
    \tilde{J}\!\left(\bm z(t), \bm u^{[p + 1]} (\cdot \,\vert\, t), \bm y_{\textnormal{d}}^\textnormal{R}\right) \leq \sum_{j = 1}^N w_j \tilde{J}\!\left(\bm z(t), \tilde{\bm u}_j (\cdot \,\vert\, t), \bm y_{\textnormal{d}}^\textnormal{R}\right) \eqqcolon \tilde J_{1:N,w}, \\
    \tilde{\bm u}_j (\cdot \,\vert\, t) \coloneqq \bm u\!\left(\bm u_{j}^{[p]} (\cdot \,\vert\, t) + \bm \gamma_j^{[p]}, \bm u^{[p]}_{-j}(\cdot \,\vert\, t) \right).
\end{align}
holds. 
If the inequality identifies the cost function as not behaving convex-like for the inserted inputs, the direction with the worst cost improvement
\begin{align}
    j_\textnormal{max} = \arg\max_j \left\{\tilde{J}\!\left(\bm z(t), \tilde{ \bm u}_j(\cdot \,\vert\, t), \bm y_{\textnormal d}^{\textnormal R}\right)\right\}
\end{align}
is eliminated.
To perform the elimination, the weight of the worst direction is set to $w_{j_\textnormal{max}} = 0$, while ensuring that the sum of the weights remains one by rescaling the weights in the form
\begin{align}
	w_j^\textnormal{new} = \frac{w_j}{\sum_{j\in\mathcal N \setminus \lbrace j_\textnormal{max}\rbrace} w_j} \quad \text{with} \quad w_{j_\textnormal{max}} = 0 \quad \forall j \in \mathcal N.
\end{align}
The algorithm then recomputes the candidate inputs~\eqref{eq:candidate_input} before checking if the cost function is convex-like.
This process is repeated until the inequality~\eqref{eq:convex_like} holds. In the worst case, \eqref{eq:convex_like} holds with one remaining direction.
When the inequality~\eqref{eq:convex_like} holds and $p < \bar p$, the algorithm continues with the first step using the current candidate input $u_{\hat\imath}$ as the initial guess for the subsequent iteration.
When the maximum number of iterations per time step is reached, the first part of the candidate input sequence $\bm u_{\textnormal{a}, \hat \imath}(t) = \bm u_{ \hat \imath}(t\,\vert\, t)$ is applied to the system and the optimization problem is solved anew in the subsequent time step with updated measurements of the state.

Since a single differential-drive robot can be asymptotically stabilized with MPC without terminal conditions~\cite{worthmann2015regulation}, neither a terminal constraint nor a terminal cost is used.
At each time step, a warm start
\begin{align}\label{eq:warm_start}
    \bm u_j^{[0]} (t + 1 + k \,\vert\, t + 1) = \begin{cases}
        \bm{u}_j^{[\bar p]} (t + 1 + k \,\vert\, t) & \text{for} \; k \in \mathbb Z_{0:H - 2}, \\
        \bm 0 & \text{for} \; k = H - 1
    \end{cases}
\end{align}
for each agent $j \in \mathcal N$ is used to initialize the values by adding a zero-input to the shifted prediction from the previous time step.
For the initialization in the first time step, we use a warm start $\bm u^{[0]}(\cdot \,\vert \,0) \neq \bm 0$ as this has been observed to help with convergence~\cite{rosenfelder2022}.
The detailed proceedings are summarized in Algorithm~\ref{alg:NonconvexOptimizer}.
\begin{algorithm}[btp]
	\caption{DMPC Nonconvex Optimizer with Autoencoder.}
	\label{alg:NonconvexOptimizer}
	\begin{algorithmic}[1]
        \STATE {\bf Input:} Initial input $\bm u^{[p = 0]}(\cdot \,\vert\, 0) \in \mathbb R^{H m_i N}$, desired input $\bm y_\textnormal{d}^\textnormal{R}$ finite $\bar p$, and $\sigma,\beta \in (0,1)$, $\alpha>0$
        \STATE At each time step $t \geq 0$: \label{inalgo:first_line}
        \begin{ALC@g}
        \STATE All robots $i \in \mathcal N$ in parallel:
        \begin{ALC@g}
            \FOR{$p\in \mathbb Z_{1:\bar p}$}
                \STATE $\alpha_i^{[p]} \leftarrow \bar \alpha$
                
                \STATE Compute $\bar{\bm u}_i^{[p]}$ using~\eqref{eq:projected_u}
                \STATE $\bm\nu_i^{[p]} \leftarrow \bar{\bm u}_i^{[p]} - \bm u_i^{[p]}$
                \WHILE{Armijo~\eqref{eq:armijo} is not fulfilled}
                    \STATE $\tilde J_i^{[p]} \leftarrow \tilde J(\bm u_i^{[p]} + \alpha_i^{[p]}\bm\nu_i^{[p]}, u_{-i}^{[p]})$
                    \STATE $\alpha_i^{[p]} \leftarrow \beta \alpha_i^{[p]}$ 
                \ENDWHILE
                \STATE Ensure $\sum_{j \in \mathcal N} w_j = 1$ with $w_j > 0 \, \forall j\ \in \mathcal N$.
                \STATE Compute candidate input $\bm u_j^{[p+1]} \leftarrow \bm u_j^{[p]} + w_j\alpha^{[p]}\bm\nu_i^{[p]}$ \label{inalgo:compute_candidate_1}
                \STATE Encode $\bm u_i^{[p+1]}$ and send code $\bm u_{i,\,\textnormal{c}}^{[p+1]}$ \label{inalgo:send}
                \STATE Receive $\bm u_{j,\,\textnormal{c}}^{[p+1]}$  and compute the reconstruction $\bm u_j^{[p+1]}\coloneqq\bm u_{j,\,\textnormal{r}}^{[p+1]}$ for each robot $j\in\mathcal N \setminus \lbrace i\rbrace$ \label{inalgo:receive}
    	        \STATE Using $\bm u_j^{[p+1]}$ compute step $\bm\gamma_j^{[p]} = \alpha_j^{[p]}\bm\nu_j^{[p]}$ for $j\in \mathcal N \setminus \lbrace i\rbrace$ with \eqref{eq:step_from_u} \label{inalgo:compute_received_candidate}
    	        \STATE $\tilde J_j^{[p]} \leftarrow \tilde J\!\left(u_j^{[p]} + \bm \gamma_j^{[p]}, u_{-j}^{[p]}\right)$ for $j\in\mathcal N\setminus \lbrace i\rbrace$
                \STATE Set $k \leftarrow 1$
                \WHILE{$k < \mathcal N$}
                    \STATE $\tilde J_{1:N,w} \leftarrow \sum_{j=1}^N w_j \tilde  J_j^{[p]}$ \label{inalgo:where_to_update_u}
                    \IF{$\bm u^{[p+1]}$ satisfies~\eqref{eq:convex_like}}
                        \STATE break
                    \ELSE
                         \STATE $j_\textnormal{max} \leftarrow \arg \max_{j\in \mathbb \mathcal N} \left\{\tilde J_j^{[p]}\right\}$
    			             \STATE $w_{j_\textnormal{max}} \leftarrow 0$
    			             \STATE $w_\textnormal{sum} \leftarrow \sum_{j \in \mathcal N} w_j$ 
    			              \FOR{ $l$ in $\mathbb Z_{1:\bar p}$}
    				            \STATE $w_l \leftarrow w_l / w_\textnormal{sum}$
                                \STATE $\bm u_l^{[p+1]} \leftarrow \bm u_l^{[p]} + w_l\bm\gamma_l^{[p]}$ \label{inalgo:compute_candidate_2}
    			            \ENDFOR
                     \ENDIF
                \ENDWHILE
            \ENDFOR
            \STATE Apply candidate input $\bm u_{a,i} = \bm u_i^{[\bar p]}(t \,\vert\, t)$ and update state measurement $\bm z_i(t+1)$
            \STATE Set $\bm u_i^{[p=0]}(\cdot \,\vert\, t + 1)$ with~\eqref{eq:warm_start}
            \STATE $t \leftarrow t + 1$ and go to~\ref{inalgo:first_line}
        \end{ALC@g}
    \end{ALC@g}
	\end{algorithmic}
\end{algorithm}
When it is not explicitly required, we refrain from denoting the current time for readability.
Note that, due to the limited calculation time available per time step in a real-time application, the number of iterations $\bar p$ is finite, and, thus, the result of the optimization is always an approximation, resulting in a suboptimal distributed model predictive controller. 
Nonetheless, convergence to an accumulation point of the overall objective function is guaranteed also for a finite number of iterations per time step~\cite{stewart2011}.
Importantly, the time available for each iteration need not only account for calculation time but also for the time needed to exchange data. 

\section{DMPC with Autoencoded Communication}\label{sec:add_ae_to_opti_algo}
Including an autoencoder into the optimization algorithm to lower the communication effort of the inter-robot communication requires multiple steps.
First, a training dataset must be created by recording the communication from sufficiently many scenarios such that the space of feasible communication samples is covered well.
Second, the autoencoder must be trained and tuned using the generated training data.
Only then can it be employed in the distributed optimization algorithm.

\begin{figure}[btp]
	\centering
    \includegraphics{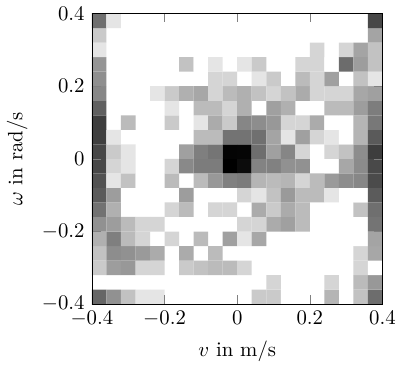}
    \includegraphics{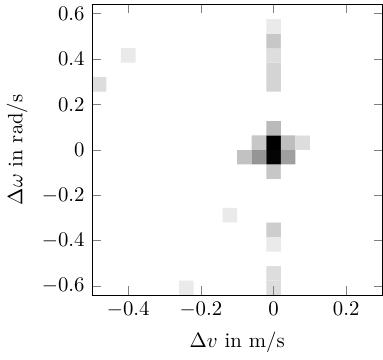}
    \caption{Heatmap comparison of the distributions of $\bm u$ and $\bm\gamma$ for the same scenarios for the the first input values in the prediction horizon corresponding to the input of one node of the autoencoder. The darker the shade of a grid cell, the more samples it contains.}\label{fig:compare_u_step}
\end{figure}

Instead of communicating the inputs $\bm u^{[p]}$ as assumed so far, it is also possible to base the communication on the step $\bm \gamma^{[p]}$, at least under the assumption of lossless communication. 
Algorithm~\ref{alg:NonconvexOptimizer} can be adapted for this by sending and receiving the steps $\bm \gamma_{\bullet}^{[p]}$ in lines~\ref{inalgo:send} and~\ref{inalgo:receive} instead of the candidate inputs.
Moreover, it is only necessary to construct $\bm u_i^{[p+1]}$ for each robot $i\in\mathcal N$ following line~\ref{inalgo:where_to_update_u} instead of in lines~\ref{inalgo:compute_candidate_1} and~\ref{inalgo:compute_candidate_2}, when communicating the step.
It could seem, at first, attractive to use the steps instead of the inputs as the to-be-compressed communication data since encoding differences to previous signals is done in many compression strategies.
However, here, there are also many intuitive reasons why communicating and applying an autoencoder to the actual (candidate) inputs instead of the steps is advantageous.
First of all, the bounds of the inputs are fixed by the constraints, so it is clear what range the training samples should cover, whereas no clear bounds are defined for the steps. 
Second, for a given initial state and setpoint, it is also much easier to estimate what values the inputs are likely to take. 
Due to the step being the weighted difference between old and new input prediction and due to MPC's receding horizon character, it is often quite small, often further decreased by the multiplication with the step size.
This intuition is further supported by Fig.~\ref{fig:compare_u_step}.
It depicts a heatmap of recorded steps and inputs at the beginning of the prediction horizon for eleven example scenarios including among others a parallel parking task.
Each heatmap is made up of 6348 samples.
The step values are clustered around zero with a few larger values. 
This combination typically requires scaling and balancing of the data before using it as training data.
The input samples are much more evenly distributed and thus more suitable.
Preliminary numerical experimentation has shown that, indeed, trying to learn an autoencoder to compress the steps yielded worse results.
Therefore, this work subsequently focuses on reducing the dimension of the inputs. 
This has the additional advantage that it does not lead to persistent errors in the internal value for other robots' candidate inputs if messages are lost. 

\begin{figure}[btp]
    \centering
    \includegraphics{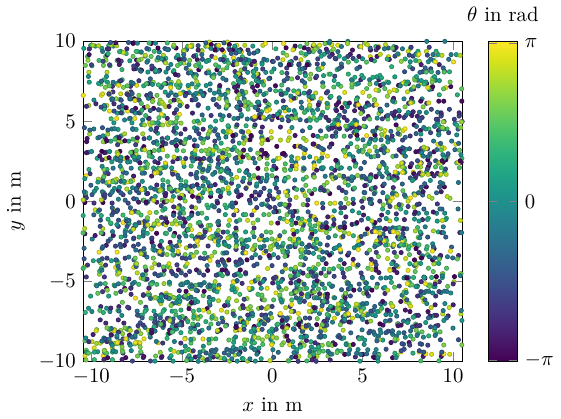}
    \caption{Initial states of the training scenarios with the color representing the initial orientation in the interval $[\qty[parse-numbers=false]{-\pi}{\radian}, \qty[parse-numbers=false]{\pi}{\radian})$}
    \label{fig:initial_positions}
\end{figure}

The training data for the autoencoder is collected by recording the complete communication between two robots for 2000 randomly generated scenarios, see Fig.~\ref{fig:initial_positions} displaying the initial position of all robots used for the training.
Each data message communicated between robots is a training sample. 
We can use certain properties of the formation-control problem and the fact that we learn a representation of candidate inputs of an OCP to make the data-collection process more efficient. 
In particular, the one required property of our sampling procedure is that the space of possible candidate inputs is covered well.
In that regard, we argue that it is sufficient to collect data from formations with two robots as we can cover well the to-be-sampled area of candidate inputs already just with two-robot scenarios. 
The results will later indicate that collecting data this way is indeed sufficient.
Secondly, in all training scenarios, the desired position of the geometric center is set to $\hat{\bm z}_\textnormal{d} = \begin{bmatrix}0\,\unit{\meter}& 0\,\unit{\meter}& 0\,\textnormal{rad}\end{bmatrix}\tran$ for each scenario.
Using the same desired state for all training scenarios is adequate as the communicated information depends on the deviation of the current state from the desired state and not on the absolute values.
Thus, the space of possible input scenarios can be covered well by considering a single desired state but different initial conditions. 
For the training scenarios, the initial position of the virtual leader $\hat{\bm z}_\textnormal{init} = \begin{bmatrix}\hat x_\textnormal{init} & \hat y_\textnormal{init} & \hat \theta_\textnormal{init}\end{bmatrix}\tran$ is chosen randomly in the ranges $\hat x_\textnormal{init} \in [\qty{-10}{\meter}, \qty{10}{\meter})$, $\hat y_\textnormal{init} \in [\qty{-10}{\meter}, \qty{10}{\meter})$, and~$\hat \theta_\textnormal{init} \in [\qty[parse-numbers=false]{-\pi}{\radian}, \qty[parse-numbers=false]{\pi}{\radian})$. 
To simulate imperfect initial placement of the robots, here and in the following, Gaussian noise is added with zero mean and standard deviations $\qty{0.0001}{\meter}$ and $\qty{0.0001}{\radian}$ to both initial position and orientation of each robot around its desired relative placement in the formation. 
The inputs are constrained to~$v \in [\qty{-0.4}{\meter\per\second}, \qty{0.4}{\meter\per\second}]$ and~$\omega \in [\qty[parse-numbers=false]{-\pi / 8}{\radian\per\second}, \qty[parse-numbers=false]{\pi / 8}{\radian\per\second}]$.
This choice of input constraints holds for all subsequent experiments.

\begin{figure}[btp]
\centering
   \includegraphics{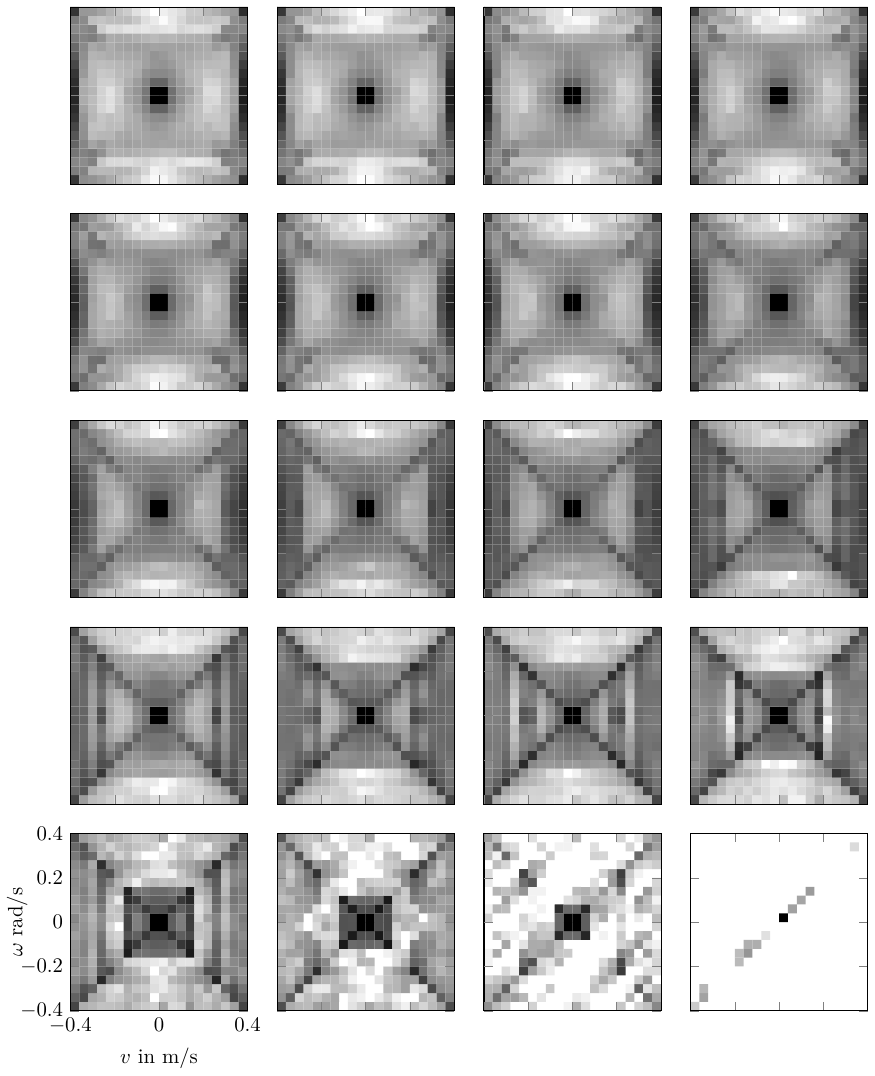}
    \caption{Heatmap of trainings data distribution over the horizon, where time steps in the prediction horizon increase from left to right and from top to bottom. Thus, the current input is depicted in the top left and the end of the prediction horizon in the bottom right corner. This corresponds to the input nodes of the autoencoder.}\label{fig:training_heatmap}
\end{figure}

For the application of an autoencoder in both simulated and real-world formation tasks, we argue that it is sufficient to record the communication of simulated scenarios.
This is the case because the optimization is based only on the current state of the robot and the remaining deviation from the setpoint.
Additionally, the optimization is influenced by the initial guess when suboptimal inputs are communicated, as is the case here.
However, the autoencoder is trained solely on the current input, and as long as the possible input values are covered well, this dependence does not impact the training. 
Unlike in many other applications of machine-learning in engineering and control, it is not (potentially noisy and faulty) measurement data that drives the training process, but aggregated, individual optimal solutions of optimization problems, which, unlike measurements, are not expected to look differently between simulations and hardware experiments. 
Figure~\ref{fig:training_heatmap} displays a heatmap of the distribution of inputs over the prediction horizon of the complete training data.

The parameters defined in the following also hold in subsequent sections where applicable and if not stated otherwise.
For all experiments, we use an initial step size of $\bar \alpha = 0.1$, the backtracking factor $\beta = 0.5$, and the Armijo factor $\sigma = 0.5$, as these parameter choices  always worked well already in previous works.
We use CasADi's auto-differentiation functionality~\cite{Andersson2019} to compute the derivative of the cost function.
To provide the formation with sufficient time to move close to the steady state without often lingering there for many time steps, the duration of the simulation is set to $T = \qty{40}{\second}$ for the training-data generation.
The sampling time is set to $\delta = \qty{0.25}{\second}$.
Computational experiments with different combinations of optimization iterations per time-step and prediction horizon led to the choice of $\bar p = 3$ and the prediction horizon length $H = 20$. 

As mentioned in Section~\ref{sec:problem_setup}, we use a separate simulator program to simulate the robots' motions.
After completing the maximum number of iterations $\bar p$, each robot $\hat \imath$ publishes its first input values $\bm u_{\hat \imath, \textnormal{a}}(t) = \bm u_{\hat \imath}^{[\bar p]}(t \,\vert\, t)$ along the prediction horizon.
The simulator integrates the system dynamics with the applied input $\bm u_{\textnormal{a}}(t\,\vert\, t)$ using a zero-order hold on the input over the duration of a sampling interval, simulating the robots' motions.
Finally, the simulator publishes the new poses of the robots $\bm z (t + 1)$ to the network, imitating a pose-tracking system. 
In our previous work, this setup allowed moving to robot hardware without changes to the control code, just having a measurement system publishing the states, and having the robotic hardware listen to the control inputs and applying them appropriately to the motors~\cite{rosenfelder2022, ebel2022cooperative}.

As in~\cite{rosenfelder2022}, we want to treat all robots equally and choose the weights in the cost function accordingly, i.e., $d_{i,j}=q_i$, $i \in \mathbb Z_{1:3}$ for $j \in\mathcal N$.
The virtual leader is treated like an additional robot such that $\hat d_i = q_i$.
The weights are chosen analogously to~\cite{rosenfelder2022} for each robot $j\in \mathcal N$ as $\hat d_1 = d_{1,j} = 1$, $\hat d_2 = d_{2,j} = 5$, $\hat d_3 = d_{3,j} = 0.1$, $r_1 = r_{1,j} = 0.125$, and $r_2 = r_{2,j} = 0.0125$ as motivated by the findings in~\cite{worthmann2015model}.
Within the solver, the cost function is scaled by a factor of $10^7$ to increase numerical accuracy. 
To compute subsequent predicted states in the cost function, the forward Euler method is applied. 

\begin{figure}[btp]
	\centering
    \includegraphics{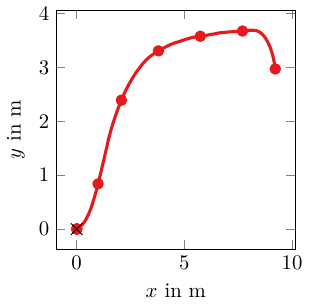}
    \includegraphics{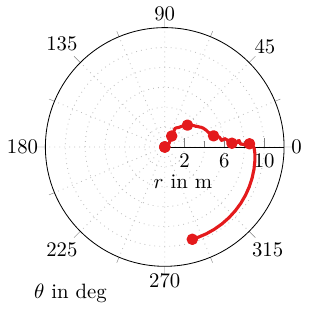}    
    \caption{Exemplary training scenario trajectory of the geometric center}\label{fig:training_trajectory}
\end{figure}

If not stated otherwise, the same model is used for simulating the multi-robot system and for predictions in the controller, i.e., there is no plant-model mismatch, apart from using error-controlled time integration for simulation.
This will allow us to study individually the error made through communication reduction.
Later, we will also look at simulations where the simulation model is considerably more realistic and complicated than the prediction model used in the controller. 
Figure~\ref{fig:training_trajectory} displays the trajectory of the geometric center of a formation in an exemplary training scenario.

We randomly split the training data into \qty{70}{\percent} used for training, \qty{10}{\percent} for validation, and \qty{20}{\percent} for the test set. For the implementation of the autoencoder with PyTorch~\cite{Ansel_PyTorch_2_Faster_2024}, we use the ADAM optimizer~\cite{AdamKingmaBa14} for stochastic optimization together with the mean squared error loss~(MSE).
Prior to training the network's parameters (its weights and biases) are initialized with the uniform distribution complying to the default behavior in PyTorch.
The input and output sizes are $m_i H = 40$ as we use a model of the system dynamics with $m_i = 2$ inputs.

\begin{figure}[btp]
	\centering
	\includegraphics{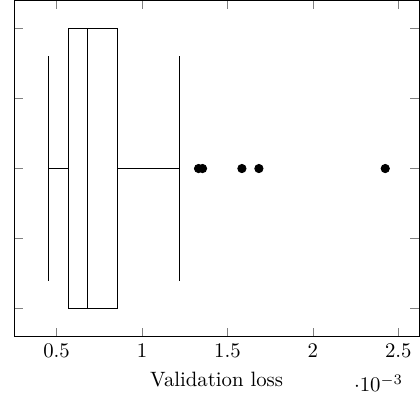}
	\caption{Distribution of lowest epoch validation loss per parameter setting. Results with code size 3 are excluded.}
	\label{fig:comparison_sorted_unsorted}
\end{figure}

A hyperparameter sweep with \num{159} runs was tracked using Weights \& Biases~\cite{wandb}.
The hyperparameter sweep led to the lowest validation loss with the parameters summarized in Table~\ref{tab:besthyperparameter}.
In the following, this hyperparameter setup is used.
\begin{table}[btp]
    \centering
    \begin{tabular}{c c c c c c c c c c}\toprule
         code l. & en (o) & en (i) & de (i) & de (o) & $n_{\text{e}}$ & a.\ f. & batch & l.\ r. & val.\ loss \\\midrule
         10 & 30 & 25 & 20 & 35 & 200 & tanh & 256 & 0.001482 & 0.0004494 \\\bottomrule
    \end{tabular}
    \caption{Hyperparameter setting with lowest mean validation loss~(val.\ loss) over an epoch. Encoder and decoder layer sizes are abbreviated with en and de, respectively, thereby differentiating inner~(i) and outer~(o) layers. The dimensionality of the reduced representation is indicated with code length (code l.). The number of epochs is $n_{\text{e}}$, a.\ f. the activation function, and l.\ r. the learning rate.}
    \label{tab:besthyperparameter}
\end{table}
The crucial parameter is the code length describing the length of the compressed representation of the communicated data.
The communicated data is communicated as an array of doubles.
Here, the original length of the communicated data is \num{40}.
Thus, compressing the length of the data to \num{10} (the chosen code length) using the autoencoder corresponds to a reduction of the communicated data by \qty{75}{\percent}. 
The different settings considered for each hyperparameter are summarized in Table~\ref{tab:consideredhyperparameter}.
\begin{table}[btp]
    \centering
    \begin{tabular}{c c }\toprule
         Parameter & Values \\\midrule
          outer encoder layer size & 26, 30, 35 \\
          inner encoder layer size & 15, 20, 25 \\
          code length & 3, 5, 8, 10\\
          inner decoder layer size & 15, 20, 25 \\
          outer decoder layer size & 25, 30, 35\\
          epochs & 30, 200, 400\\
          activation Function & sigmoid, ReLU, leaky ReLU, tanh \\
          batch size & 256, 512 \\
          learning rate &  $\textnormal{round}(X/q)q$, $q=1\cdot 10^{-6}$ and $X\in (0, 0.01]$ uniform  \\
          optimizer & ADAM, SGD \\\bottomrule
    \end{tabular}
    \caption{Summary of considered hyperparameters}
    \label{tab:consideredhyperparameter}
\end{table}
The last layer applies a linear activation function in all cases independently of the nonlinear activation function.
Apart from the code length \num{3}, which is dysfunctional, it seems the performance is relatively robust towards changes in the hyperparameter setup, indicating that our learning task seems comparatively well posed, see Fig.~\ref{fig:comparison_sorted_unsorted}.

\section{Results with Autoencoder-Based Communication Reduction}\label{sec:results}
In this section, results are presented comparing the performance of the robot formations driving to static setpoints with and without reduced communication.
We first present results in an idealized setting without plant-model mismatch before moving on to a more realistic setup with plant-model mismatch.
Finally, the communication reduction is applied in a distributed setup with each suboptimization problem being solved on a separate, wirelessly communicating single-board computer typical for mobile robots instead of on a single, powerful machine.

\subsection{Simulative Analysis}\label{sec:sim_results}
To compare the behavior of robot formations with full-message communication with reduced-message communication in an otherwise idealized setting, \num{200} randomized test scenarios are considered.
This corresponds to a tenth of the scenarios used to record the training data.
The \num{200} scenarios are equally split up into formations with $N\in\mathbb Z_{2:6}$ robots, i.e., \num{40} formations are considered per number of robots.
Each scenario is simulated once without communication reduction and once with communication reduction using an autoencoder.
When we speak of communication reduction or full and reduced communication, we refer purely to the size of messages rather than to the quantity of messages. 
Besides comparing the performance with and without communication reduction, we want to test our assumptions that formations with two robots are enough to collect data for training and that it suffices to drive the formation center to $\hat{\bm z}_\textnormal{d} = \begin{bmatrix}\qty{0}{\meter}& \qty{0}{\meter}& \qty{0}{\radian}\end{bmatrix}\tran$.
Additionally, we want to investigate the performance on a domain that is larger than the training domain, to see how the formations behave when the distances to the desired output are increased.
Thereby, we check our previous suggestion that we covered the sampling space well enough.
For these reasons, each test task starts with the random initial pose of the geometric center in $x,y\in [\qty{-20}{\meter},\qty{20}{\meter})$  and $\theta\in[\qty[parse-numbers=false]{-\pi}{\radian},\qty[parse-numbers=false]{\pi}{\radian})$.
The random numbers are uniformly distributed.
Different from the training, the desired pose of the geometric center is also chosen randomly within the same region as the initial pose.

The use of a larger area and formations consisting of more robots requires the use of a larger time period to be able to empirically check for convergence within the simulation duration. 
A simulation duration of $T=\qty{200}{\second}$ is used for formations of two to four, $\qty{300}{\second}$ for five, and $\qty{450}{\second}$ for six robots.
Apart from this, a parameter setup identical to the training setup is used for the optimization algorithm and combined with the best performing hyperparameters for the autoencoder as described in Section~\ref{sec:add_ae_to_opti_algo}.

\begin{figure}[btp]
    \centering
    \includegraphics{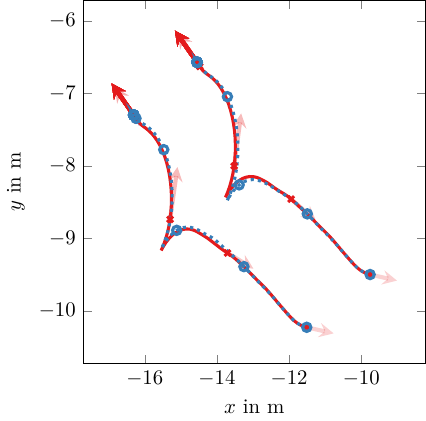}
    \includegraphics{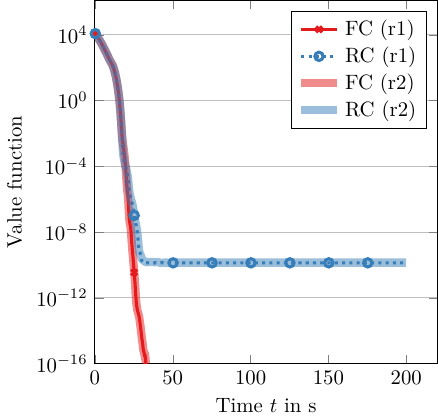}
    \caption{Left: The closed-loop trajectories of a robot formation with two robots with and without reduced communication. The arrows in the left plot indicate the robot orientations for the task with full communication. Right: The corresponding cost function.}
    \label{fig:exemplarytask}
\end{figure}

\begin{figure}[btp]
	\centering
    \includegraphics{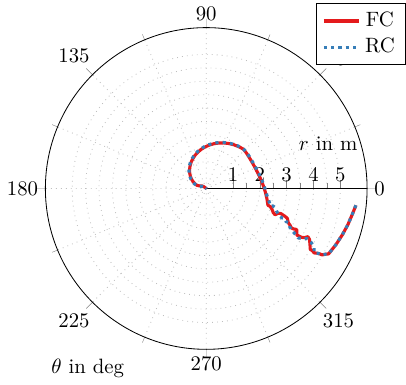}
    \caption{Orientation comparison of the exemplary test scenario using the geometric center.}\label{fig:test_orientations}
\end{figure}

Figure~\ref{fig:exemplarytask} displays the performance of the DMPC controller with full and reduced communication of candidate inputs in one of the example scenarios.
Full communication~(FC) is depicted by a red line~(\tikz{\draw[mycolor1, line width=1.5pt] (0,0) -- (0.5,0);}).
The arrows indicate the orientation of the robots and the opacity increases with time, i.e., each subsequent time step is more opaque than the previous.
When the communication is reduced (RC), the robots' paths are depicted by a blue line~(\tikz{\draw[mycolor2, dotted, line width=1.5pt] (0,0) -- (0.5,0);}).
To keep the figure comprehensible, the orientation indication is omitted for the case with reduced communication. 
Instead, the orientations of the virtual leaders are compared in Fig.~\ref{fig:test_orientations} showing that also the orientations follow similar trajectories.
Therein, the $\theta$-axis corresponds to the orientation and the $r$-axis is the distance to the desired position of the virtual leader.
In the depicted task, the initial states of the robots are set to
\begin{align}
    \bm z(t_0) &\coloneqq \begin{bmatrix}
        \qty{-11.51}{\meter} & \qty{-10.22}{\meter} & \qty{-0.11}{\radian} & \qty{-9.76}{\meter} & \qty{-9.50}{\meter} & \qty{-0.11}{\radian}
    \end{bmatrix}\tran
\intertext{and the desired output to}
    \bm y_\textnormal{d}^\textnormal{R} &\coloneqq \begin{bmatrix} \hat{\bm z}_\textnormal{d} & \bm z_{\hat{\bm z}\to 1,\textnormal{d}}^\textnormal{R} &
 \bm z_{\hat{\bm z}\to 2, \textnormal{d}}^\textnormal{R}
\end{bmatrix}\tran
\end{align}
with 
\begin{align}
    \hat{\bm z}_\textnormal{d} &= \begin{bmatrix}
        \qty{-15.43}{\meter} & \qty{-6.93}{\meter} & \qty{2.52}{\radian}
    \end{bmatrix}\tran, \; {\bm z}_{\hat{\bm z}\to 1, \textnormal{d}}^\textnormal{R} = \begin{bmatrix}
        \qty{-0.88}{\meter} & \qty{-0.36}{\meter} & \qty{2.52}{\radian}
    \end{bmatrix}\tran,\\ 
    \;
{\bm z}_{\hat{\bm z}\to 2, \textnormal{d}}^\textnormal{R}  &=\begin{bmatrix}
        \qty{0.88}{\meter} & \qty{0.36}{\meter} & \qty{2.52}{\radian}
    \end{bmatrix}\tran.
\end{align} 
The values in both vectors are rounded to two decimal places for readability.
One can observe that the two cases follow similar trajectories regarding both position and orientation.
A look at the optimal value function on the right of Fig.~\ref{fig:exemplarytask} shows that the full communication outperforms the reduced communication by several orders of magnitude towards the end of the formation task, although the reduced communication still achieves satisfactory accuracy levels, better than the accuracy of typical position measurements would be.
The overall cost function is always computed using the information from one specific robot with the inputs received from the other robots, which means that errors made by the autoencoder transition into the cost function.
To highlight that the cost function takes a similar value on each robot, the cost function of the other robot is depicted with lower opacity.

Looking at the average, best, and worst cost function values at the end of each test task supports this observation further, see~Table~\ref{tab:finalcostcomparison}.
\begin{table}[btp]
	\centering
	\begin{tabular}{c c c c}\toprule
		Type & average cost & lowest cost & highest cost \\\midrule
		FC &  $2.5709\cdot 10^{-15}$ & $<10^{-16}$ & $1.964\cdot 10^{-13}$ \\
		RC & $6.7161 \cdot 10^{-10}$ & $1.3425 \cdot10^{-10}$ & $4.3815 \cdot 10^{-9}$ \\\bottomrule
	\end{tabular}
	\caption{Comparison of the optimal cost values at the final time of each of the \num{200} test tasks with two to six robots.}
	\label{tab:finalcostcomparison}
\end{table}
With an average final cost of less than $1\cdot10^{-9}$, the communication reduction performs sufficiently well to drive a robot formation close to the setpoint.
In the beginning, when the cost is still high, the evolution of the cost function values aligns well.
This alignment indicates that a formation with reduced communication manages to drive close to the desired output within approximately the same time as formations with full communication.

\begin{figure}[btp]
    \centering
    \includegraphics{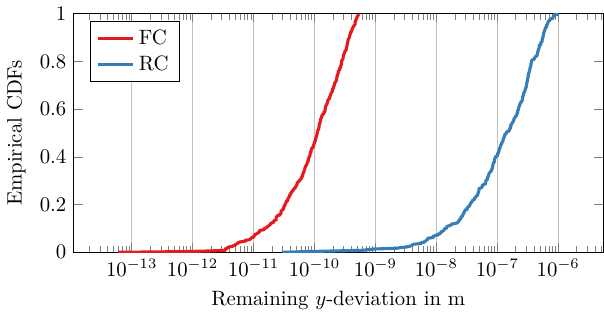}
    \caption{Empirical cumulative distribution functions~(CDFs) of remaining deviations in the hard-to-control $y$-direction for \num{120} random formation tasks with full and autoencoder-reduced communication. The data is collected from scenarios with two to four robots at time $t=\qty{300}{\second}$.}
    \label{fig:cfd_y_direction}
\end{figure}

To specifically investigate the terminal behavior of the robot formations, we consider another \num{120} random formation scenarios on a region of $x,y\in[\qty{-2}{\meter},\qty{2}{\meter})$ with two to four robots with a desired orientation of $\hat \theta_\textnormal{d} = \qty{0.0}{\radian}$. 
The simulation duration is set to the long duration of $T = \qty{300}{\s}$ to, in conjunction with the initial conditions, give (more than) enough time for practically complete convergence with the given number of robots in the given area so that we can be sure that left-over deviations at the end of the simulation are not just a product of incomplete convergence due to too little convergence time given. 
We focus the discussion on the $y$-direction as, for the uniform setting of $\hat \theta_\textnormal{d} = \qty{0.0}{\radian}$, it is the direction that is, terminally, hardest to control given the nonholonomic differential-drive robot from Section~\ref{sec:theory}, cp.~\cite{rosenfelder2022,rosenfelder2023automatica}.
Figure~\ref{fig:cfd_y_direction} depicts the resulting empirical cumulative distribution functions~(CDFs) of the deviation in the crucial $y$-direction at $t=\qty{300}{\second}$ providing statistics on how closely the robots manage to approach their respective desired state in the hardest-to-control $y$-direction. 
Each robot's deviation enters the CDF, but the virtual leader is excluded. 
Whereas the case with full communication achieves a significantly lower remaining deviation, the variant with reduced communication still achieves accuracies far below one millimeter reliably. 
In contrast, in a real-world application, the robots are anyway not expected to be able to approach the setpoint with an accuracy of less than a millimeter as typical inaccuracies through imperfect hardware and measurements are often larger than the approximation error obtained here for reduced communication. 
Therefore, in real-world applications, the terminal deviation introduced by the approximation made by the autoencoder will typically be dominated by other sources of error, as even typical position-measurement errors of high-performance motion-tracking systems for laboratory usage are larger. 

\subsection{Simulation with Plant-Model Mismatch}\label{sec:results_plant_model_mismatch}
In this section, we consider a scenario with a plant-model mismatch to make the simulation more realistic by using a significantly more complex model in the simulator.
Therefore, a multi-body model of the robot, taking into account actuation dynamics and inertia of chassis and wheels, is considered. 
The model has been successfully applied in previous work~\cite{ebel2022finding,ebel2024coop_ob_trans}, where it has shown good transferability from simulation to hardware. 
The multi-body model is given by 
\begin{align}\label{eq:multi_body_model}
    {\bm f}_i = \begin{bmatrix}
        \dot\varphi_{\textnormal{l}, i} \\ \dot\varphi_{\textnormal{r}, i} \\ \dot \omega_{\textnormal{l}, i} \\ \dot\omega_{\textnormal{r}, i} \\ \dot\theta_i \\ \dot x_i \\ \dot y_i
    \end{bmatrix}= \begin{bmatrix}
        \omega_{\textnormal{l},i} \\ 
        \omega_{\textnormal{r}, i} \\ 
        1/(2d_1d_3)\left((d_1d_4+d_2d_3)M_{\text{l},i}+(d_1d_4-d_2d_3)M_{\text{r},i}\right) \\     
        1/(2d_1d_3)\left((d_1d_4-d_2d_3)M_{\text{l},i}+(d_1d_4+d_2d_3)M_{\text{r},i}\right)\\ 
        \left(\omega_{\textnormal{r}, i}(t)-\omega_{\textnormal{l},i}(t)\right)r_\textnormal{w}/(2r_\textnormal{kin}) \\
        {}^\textnormal{R}{v_{\textnormal{C}_\textnormal{x},i}}(t)\cos(\theta(t)) \\ 
        {}^\textnormal{R}{v_{\textnormal{C}_\textnormal{x},i}}(t)\sin(\theta(t))
    \end{bmatrix}
\end{align}
with the angles $\varphi_{\textnormal{l}, i}$ and $\varphi_{\textnormal{r}, i}$ describing the orientations of the left and the right wheel, respectively, the angular velocities  $\omega_{\textnormal{l},i}$ and $\omega_{\textnormal{r}, i}$ of the left and the right wheel, as well as the left and right motor moments $M_{\text{l},i}$ and $M_{\text{r},i}$.
The constants $r_\textnormal{kin}$ and $r_\textnormal{w}$ correspond to the kinematic radius and the wheel radius, respectively.
The kinematic radius is the distance from the center of the chassis to the projection of the wheel's contact point, i.e., half the distance between the wheels.
The constants $d_1$ to $d_4$ are defined as 
\begin{align}
    d_1&\coloneqq I_\textnormal{c} r_\textnormal{w}/(2 r_\textnormal{kin})+I_{x}r_\textnormal{kin}/r_\textnormal{w}+r_\textnormal{kin}r_\textnormal{w}m_\textnormal{r}+Ir_\textnormal{w}/r_\textnormal{kin},\\
    d_2&\coloneqq r_\textnormal{kin}/r_\textnormal{w},\\
    d_3&\coloneqq (r_\textnormal{w}/2)(m_\textnormal{c}+2m_\textnormal{r}+2/(r_\textnormal{w}^2)I_{x}),\\
    d_4&\coloneqq 1/r_\textnormal{w},
\end{align}
with the moment of inertia~$I_\textnormal{c}$ of the chassis, one wheel's moment of inertia $I_{x}$ about the $x$-axis of the wheel coordinate system, and the moment of inertia~$I$ about the other relevant principle axis of the wheel.
Furthermore, $m_\textnormal{r}$ corresponds to the mass of one wheel and $m_\textnormal{c}$ to the mass of the robot chassis.
The variables $x_i$, $y_i$, and $\theta_i$ correspond to the position and orientation of the chassis.
The left and right wheel velocities can be computed from the linear velocity $v_i$ and angular velocity $\omega_i$ of the robot as $\omega_{\textnormal{l},i} = (v_i - r_\textnormal{kin}\omega_i)/r_\textnormal{w}$ and $\omega_{\textnormal{r},i} = (v_i + r_\textnormal{kin}\omega_i)/r_\textnormal{w}$.
Furthermore, it holds that ${}^R{v_{\textnormal{C}_{x},i}}(t)=  (\omega_{\textnormal{r}, i}(t)-\omega_{\textnormal{l},i}(t))r_\textnormal{w}/2$.
The simulations use the values $m_\textnormal{c}=\qty{1.73}{\kg}$, $r_\textnormal{kin}=\qty{0.12}{\m}$, $r_\textnormal{w}=\qty{0.035}{\m}$, $m_\textnormal{r}=\qty{0.0368}{\kg}$, $I_\textnormal{c}=\qty{0.01814878}{\kg\square\m}$, $I_{x}=2.254\cdot10^{-5}\,\unit{\kg\square\m}$, and $I=1.1466\cdot 10^{-5}\,\unit{\kg\square\m}$.
This model was used in previous research on the formation control of differential-drive robots in~\cite{ebel2022finding} and it turned out to be accurate enough to permit a direct transfer of methods developed in simulations to physical hardware experiments in~\cite{ebel2024coop_ob_trans}.
A simulator program simulates the movement of each robot $i\in \mathcal N$ by solving the differential equation $\dot{\bm z}_{\textnormal{sim},i}(t+1) = \bm f(\bm z_{\textnormal{sim},i}(t))$ with $\bm z_{\textnormal{sim},i} = \begin{bmatrix}
    \varphi_{\textnormal{l},i} & \varphi_{\textnormal{r},i} & \omega_{\textnormal{l},i} & \omega_{\textnormal{r},i} & \theta_i & x_i & y_i 
\end{bmatrix}\tran$, coupled with two independent PI controllers governing the motor moments to reach prescribed desired angular wheel velocities. 
The simulation uses error-controlled time integration.

\begin{figure}[btp]
    \centering
    \includegraphics{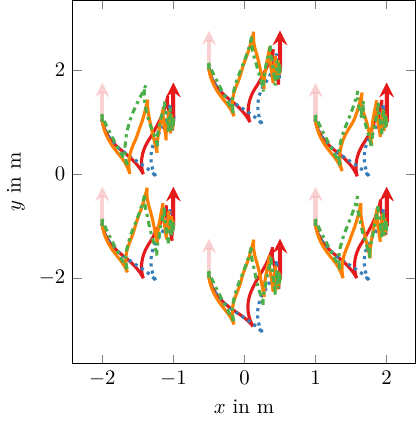}
    \includegraphics{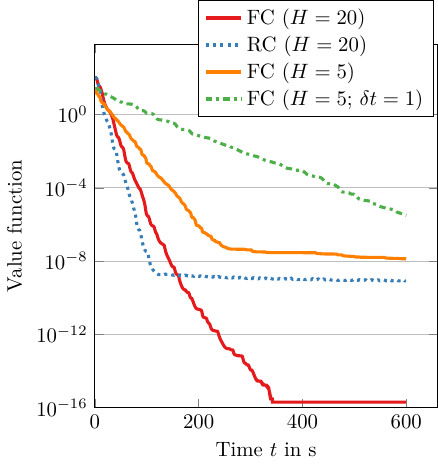}
    \caption{Left: The closed-loop trajectories of a parallel parking task with six robots for the four considered cases. The arrows indicate the orientation of the task with full communication. Right: The corresponding value function.}
    \label{fig:exem_taskparapark6}
\end{figure}

As an exemplary scenario with plant-model mismatch, Fig.~\ref{fig:exem_taskparapark6} depicts a parallel parking task with six robots.
The initial and goal states of the geometric center are $\hat x_\textnormal{d} = -\hat x(t_0)= \qty{0.5}{\m}$, $\hat y_\textnormal{d} = \hat y(t_0)= \qty{0.0}{\m}$ and $\hat \theta_\textnormal{d} = \hat \theta(t_0)= \qty[parse-numbers=false]{\pi/2}{\radian}$.
We use a simulation duration of $T = \qty{600}{\s}$.
Note that, as $\hat\theta_\textnormal{d} = \qty[parse-numbers=false]{\pi / 2}{\radian}$, the $x^\textnormal{R}$-axis of the reference frame of the chassis corresponds to the $y$-axis of the inertial frame of reference.
The reduced communication converges to a cost function value matching the observations in the simulative experiments without plant-model mismatch.
Moreover, although the trajectories obtained when using the autoencoder are clearly distinct from the corresponding results with full communication, the formation moves close the desired setpoint in about the same time~--~right until the autoencoder-based variant has reached its accuracy limit.
As the literature seemingly does not offer comparable results or methodology, the only obvious comparisons we can conduct is to a more naive communication reduction method.
One naive way is to simply reduce the prediction horizon.
As the orange line~(\tikz{\draw[mycolor4, line width=1.5pt] (0,0) -- (0.5,0);}) shows, the robot formation requires significantly more time to drive close to the desired state when the prediction horizon is set to $H=5$, assuming the same sampling time. 
With this shortened prediction horizon, the published horizon information has the same dimension as the autoencoded larger prediction horizon.
When the sampling time is increased to $\delta t = \qty{1}{\second}$, such that the $H=5$ prediction steps cover the same time interval as the original prediction horizon, the performance becomes even worse, as shown by the gray line~(\tikz{\draw[mycolor6, dashdotted, line width=1.5pt] (0,0) -- (0.5,0);}).
\begin{figure}[btp]
    \centering
    \includegraphics{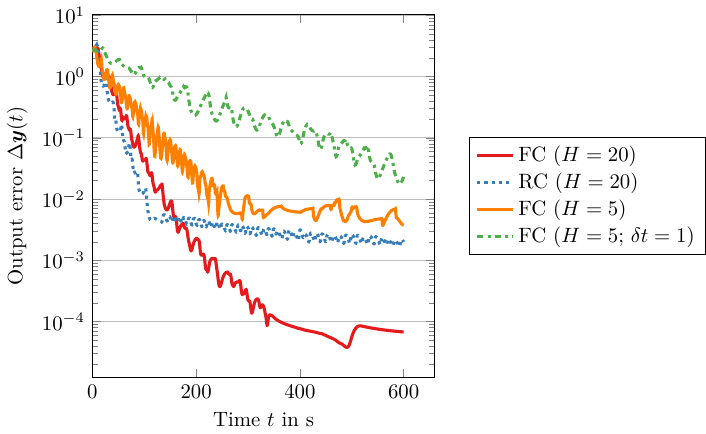}
    \caption{Comparison of the output error over time}
    \label{fig:exem_taskparapark6_y_err}
\end{figure}
Figure~\ref{fig:exem_taskparapark6_y_err}, displaying the overall output error $\Delta \bm y (t) \coloneqq \left\Vert \bm y^\textnormal{R} (\bm z(t), \hat \theta_\textnormal{d}) - \bm y_\textnormal{d}^\textnormal{R}\right\Vert_2$ over time in the Euclidean norm supports this observation further.
We, thus, observe that a reduction of the communicated data using an autoencoder works well also in a more realistic simulation and outperforms a naive shortening of the prediction horizon. 
It remains, however, to be seen if using the proposed autoencoder-based setup can actually be advantageous in physically distributed scenarios. 
 
\subsection{Numerical Experiments on Embedded Hardware with Communication and Physically Distributed Computation}\label{sec:experimental_results}
\begin{figure}
    \centering
    \includegraphics[width=0.5\linewidth]{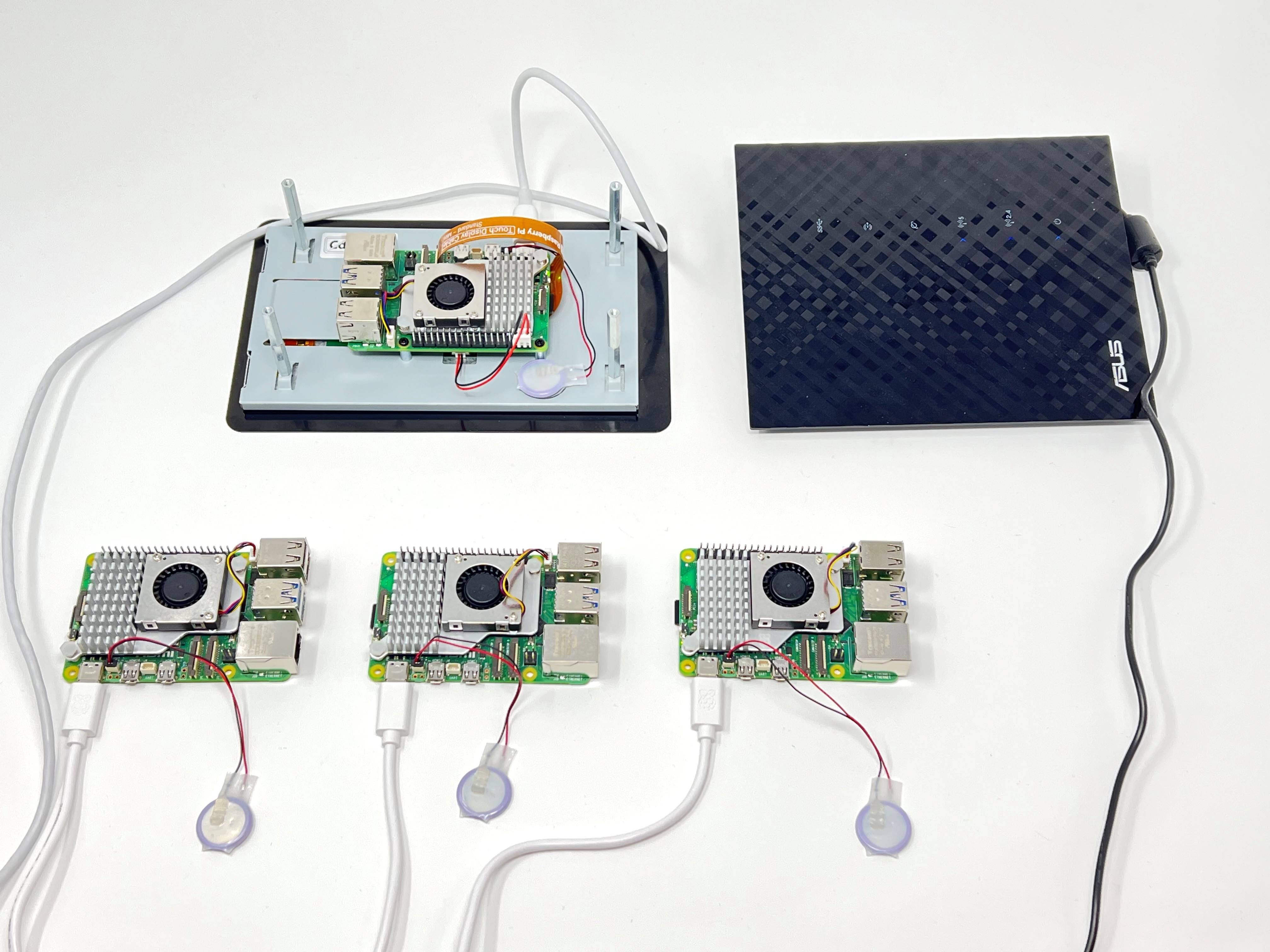} 
    \caption{Photograph of the four Raspberry Pi single-board computers and the Wi-Fi router used for the experiments on embedded hardware. One single-board computer is attached to a display (face down in the photograph) for easier debugging during development. }
    \label{fig:Embedded_setup}
\end{figure}

\begin{figure}
 \includegraphics{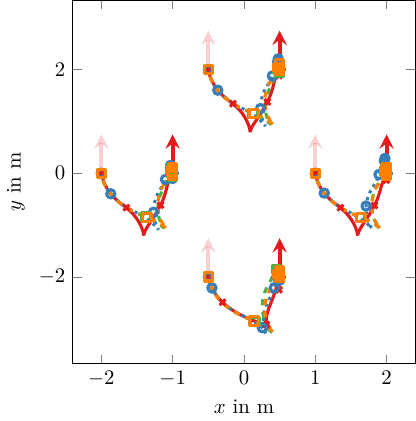}
 \includegraphics{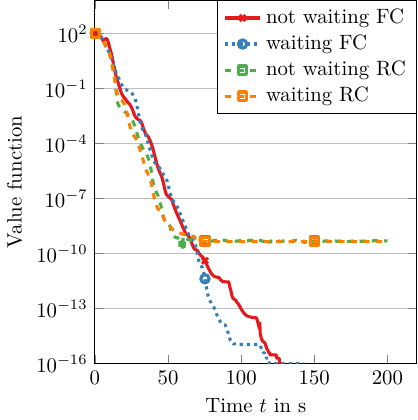}
 \caption{Left: The closed-loop trajectories of a parallel parking task with four robots for the four
considered cases. Right: The corresponding cost function.}\label{fig:embedderesults}
\end{figure}

Using the distributed simulation setting from the previous section, this section is dedicated to numerical experiments on embedded hardware with physically distributed computation. 
The setup consists of four Raspberry Pi 5 single-board computers with 8 GB RAM as they can be mounted to mobile robots. 
For instance, the mobile robot from Figure~\ref{fig:mobilerobot} uses this single-board computer model. 
Meanwhile, the simulator is run on an external personal computer, with an Intel Core i7-8550U @ 1.80 GHz processor and 16 GB RAM, and uses the more realistic multibody dynamic model. 
The hardware employed is depicted in Fig.~\ref{fig:Embedded_setup}.
All computers communicate via 2.4 GHz Wi-Fi. 
We consider a parallel parking task with four robots and a simulation duration of $T=\qty{200}{\s}$.
Clearly, in real-world experiments with physically distributed computation and lossy wireless communication, performance can vary, e.g., as network conditions like the crowdedness of the electromagnetic spectrum can vary. 
We thus conduct a few repetitions of each numerical experiment in this section to get an impression whether performance is consistent. 
The experiments were conducted in a building of LUT University with typically many wireless devices of staff and students present. 
Figure~\ref{fig:embedderesults} shows exemplary results from a parallel parking task with $\bar{p}=3$ iterations per time step as in previous sections. 
Generally, the results on embedded hardware agree well with the networked simulation results from the previous sections, with the value functions reaching the same orders of magnitude in the respective setups as before.
In the results, two cases of handling communication are distinguished.
In one case, the robots wait indefinitely until they receive the messages they request, denoted as ``waiting''. 
While a robot waits, it periodically resends its previously sent data, to cover for potentially lost messages. 
Generally, this setup could lead to time-step overruns, i.e., non-real-time operation, when subject to message losses. 
However, it represents a useful performance baseline and it corresponds closely to the setup used previously when generating the training data and conducting the simulative analyses. 
In the other case, each robot obeys the time-step constraint and moves on even if messages were not received in time using the latest available information, denoted as ``not waiting'', which means not waiting longer than permissible for real-time execution. 
The simulator always waits until it has received data from all robots before continuing.

In either case, the robots always use the latest available information, i.e., if robot $\hat\imath$ waits for messages with number $k$ from the other robots $j\in \mathcal N \setminus\lbrace\hat\imath\rbrace$ but receives message numbered $k+1$ from one, it uses this message and reuses it in the next step.
Due to this, the receive percentage is less than~\qty{100}{\percent} even if robots wait for messages to arrive as a low number of messages is skipped even when waiting because it can happen that newer information is available first.
The percentage of received messages is greater than~\qty{98}{\percent} on average with and without waiting. 
As expected, the receiving percentage is lower when the robots do not wait until messages arrive, see Table~\ref{tab:iter3embedded}.
\begin{table}[btp]
	\centering
	\begin{tabular}{c c c}\toprule
		Type & Avg.\ number of messages received in \% & Avg.\ runtime in s \\\midrule
		FC (not waiting) & 98.67 & 199.87 \\
		FC (waiting) & 99.63 & 202.78\\
		RC (not waiting) & 98.64 & 199.83 \\
		RC (waiting) & 99.78 & 201.48 \\\bottomrule
	\end{tabular}
	\caption{Average results from five experiments on embedded hardware}
	\label{tab:iter3embedded}
\end{table}
Additionally, the average runtimes when robots wait for messages only exceed the simulation duration slightly, indicating that almost lossless communication is almost feasible in real-time. 
Therefore, we expect this setup to perform well in practical scenarios using Wi-Fi and with hard time-constraints.

To see how the performance changes under more demanding conditions, we increase the number of iterations to~$\bar{p}=10$ while keeping the sampling time the same.
As the robots exchange data in each iteration, this increase of the number of iterations shortens the time a robot can wait for a message within an iteration while still being real-time capable significantly.
Instead of a third of the total time allocated to one iteration during a time step with $\bar{p}=3$, each iteration gets only a tenth with $\bar{p}=10$, resulting in less than $\qty{0.025}{\s}$ per iteration for computation and communication.
At the same time, the amount of communication is more than tripled (not including repeated sending), thus increasing the load on the communication network considerably.

\begin{figure}[btp]
    \includegraphics{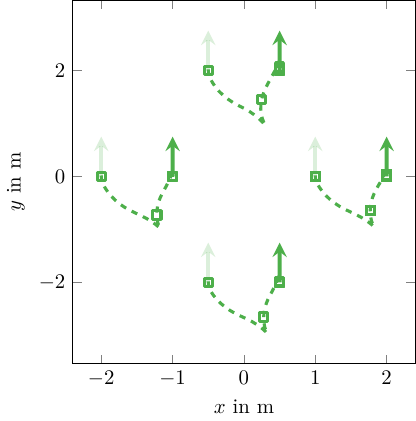}
    \includegraphics{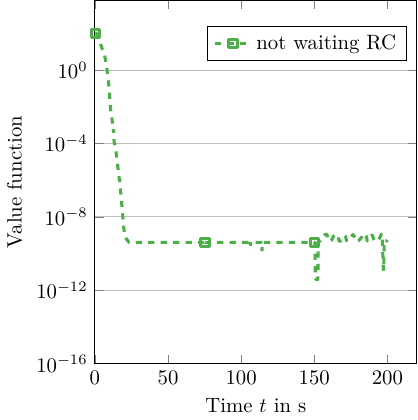}
	\caption{Left: The closed-loop trajectories of a parallel parking task with four robots for the case with reduced communication and without waiting with $\bar{p}=10$. Right: The corresponding cost function.}\label{fig:embedderesults10}
\end{figure}

Under these conditions, the experiments using full communication do not manage to exchange any messages within the sampling time of each time step, see Table~\ref{tab:iter10embedded}.
\begin{table}[btp]
	\centering
	\begin{tabular}{c c c}\toprule
		Type & Avg.\ number of messages received in \% & Avg.\ runtime in s \\\midrule
		FC (not waiting) & 0 & 200.03 \\
		FC (waiting) & 99.65 & 453.07 \\ 
		RC (not waiting) & 96.67 & 199.91 \\
		RC (waiting) & 98.33 & 205.82\\\bottomrule
	\end{tabular}
	\caption{Average results from two experiments on embedded hardware with $\bar{p}=10$}
	\label{tab:iter10embedded}
\end{table}
When the communication is reduced with the proposed autoencoder-based approach, the robots still successfully receives the desired information over \qty{95}{\percent} of the time.
Allowing the agents to wait for messages results in a computation time of more than double the simulated real-time when using full communication.
When the communication is reduced, on the other hand, the numerical experiment with waiting only takes a few more seconds than the simulated time.
In particular, the realistic setting without waiting and reduced communication leads to a functional behavior and fulfills hard real-time constraints, see the plotted trajectory and value function in Figure~\ref{fig:embedderesults10}.
The average terminal deviation in the hard-to-control $y^\textnormal{R}$-direction is $9.53 \cdot 10^{-5}\,\text{mm}$.
We expect that these results could be improved further by optimizing the communication, e.g., how often messages are re-sent when no message is received. 
However, such tuning might need to be re-done whenever network conditions change and, hence, we refrained from embarking on tuning of this kind. 
The fact that the version with reduced communication and waiting manages to solve the problem almost in the same time as the simulated time also indicates that the parts of the programs not dealing with communication indeed run fast enough.  
This shows that, in the studied scenarios, it is indeed the communication that limits the performance with full communication. 
Moreover, it has been seen that the autoencoder-based communication reduction can alleviate this problem well enough to reach real-time suitability in settings where full communication fails completely, even considering that the encoding and decoding comes with an additional (but in this study seemingly negligible) computational overhead.
\section{Conclusion}\label{sec:conclusion}
Recognizing that excessive communication is a key limiting factor for the real-world application of distributed model predictive control, in this work, we presented an approach to reducing the communication effort for distributed model predictive control by means of a message-size reduction using an autoencoder.
As an application example, the methodology was applied to formation control of differential-drive robots, whose nonholonomic constraints require nonlinear control and are challenging for optimal control in general, yielding favorable results for the proposed approach. 
To train the autoencoder for usage as a part of a distributed optimization approach, data has been generated by simulating 2000 scenarios of formations of two robots. 
It has been found that this training setup covers the space of the communicated data well, so that an autoencoder trained this way has also shown similarly good performance when considering very different scenarios than in the training data, e.g., a larger number of robots. 
Concretely, the results show that formation scenarios with a reduction of the message size in the inter-robot communication by \qty{75}{\percent} perform sufficiently well.
Crucially, we compared the proposed approach to reduced communication with a more naive approach that simply shortens the controllers' prediction horizon so that the same amount of data is communicated as with the autoencoder. 
Indeed, our proposed approach showed significantly better performance than the naive approach, hinting at the efficiency of the learned encoding.
Moreover, we have conducted experiments with physically distributed computation, Wi-Fi communication, and the optimization algorithms running on single-board computers common in robotics. 
In this realistic setting, our proposed approach even worked in scenarios in which traditional full communication failed completely, as the full data could not be communicated in a timely enough manner to meet real-time requirements.  
Future work will apply this methodology in practical mobile robotics and swarm-control applications. 

\section*{Acknowledgments}
T.\ Schiz acknowledges support from the Erasmus+ program of the European Union.


\end{document}